\documentclass[acmsmall,screen,nonacm]{acmart}

\usepackage{ifthen}
\usepackage[normalem]{ulem} 
\usepackage{xcolor}

\newboolean{showedits}
\setboolean{showedits}{true} 
\ifthenelse{\boolean{showedits}}
{
	\newcommand{\del}[1]{\textcolor{red}{\sout{#1}}} 
}{
	\newcommand{\del}[1]{} 
	
}

\newboolean{showcomments}
\setboolean{showcomments}{true}
\newcommand{\id}[1]{$-$Id: scgPaper.tex 32478 2010-04-29 09:11:32Z oscar $-$}

\ifthenelse{\boolean{showcomments}}
{\newcommand{\nbc}[3]{
 {\colorbox{#3}{\bfseries\sffamily\scriptsize\textcolor{white}{#1}}}
 {\textcolor{#3}{\sf\small$\blacktriangleright$\textit{#2}$\blacktriangleleft$}}}
 }
{\newcommand{\nbc}[3]{}
 \renewcommand{\del}[1]{} 
 }
 \definecolor{darkyellow}{RGB}{255, 222, 9}

\definecolor{ibcolor}{rgb}{0.4,0.6,0.2}
\definecolor{ascolor}{rgb}{0,0.5,0.9}

\definecolor{bocolor}{rgb}{0.6,0.9,0.2}
\definecolor{jrcolor}{rgb}{0.5,0,0.5}
\definecolor{nrcolor}{rgb}{0.4,0.1,0.3}
\definecolor{hkcolor}{rgb}{1.0,0.5,0.3}
\definecolor{tdcolor}{rgb}{1.0,0,0}

\usepackage{array}
\usepackage[utf8]{inputenc}
\usepackage{microtype}
\usepackage{hyperref}
\usepackage{graphicx}
\usepackage{xspace}
\usepackage{multicol}
\usepackage{tabularx}
\usepackage{pcallst}
\usepackage{simplebnf}
\usepackage{subcaption}
\usepackage{enumitem}
\usepackage{makecell}

\newcommand\tlaplus{TLA$^+$\xspace}

\newcommand\omnilink{OmniLink\xspace}

\newfloat{listing}{htbp}{loe}
\floatname{listing}{Listing}

\newcommand{\mpcal}[1]{\inlpluscal#1}

\definecolor{dkgreen}{rgb}{0,0.6,0}
\definecolor{orange}{rgb}{1.0,0.49,0.0}
\definecolor{bluegray}{rgb}{0.4, 0.6, 0.8}
\definecolor{dpcolor}{rgb}{0.58,0,0.82}
\definecolor{pcalcolor}{rgb}{0,0,0}  

\author{Finn Hackett}
\email{fhackett@cs.ubc.ca}

\author{Evan Wrench}
\email{ewrench@student.ubc.ca}

\author{Peter Macko}
\email{peter.macko@mongodb.com}

\author{A. Jesse Jiryu Davis}
\email{jesse@mongodb.com}

\author{Yuanhao Wei}
\email{yuanhaow@cs.ubc.ca}

\author{Ivan Beschastnikh}
\email{bestchai@cs.ubc.ca}

\makeatletter
\newcommand{\thickhline}{%
    \noalign {\ifnum 0=`}\fi \hrule height 1pt
    \futurelet \reserved@a \@xhline
}
\newcolumntype{I}{@{\hskip\tabcolsep\vrule width 1pt\hskip\tabcolsep}}
\makeatother

\begin{document}

\title{\omnilink: Specification-Based Trace Validation of Concurrent Systems with Substep Linearizability}

\begin{abstract}

Concurrent systems are notoriously difficult to validate.
We present \textbf{\omnilink}, a methodology for validating traces of
concurrent implementations against high-level specifications in
\tlaplus. Unlike prior \tlaplus based approaches,
\omnilink treats system events as black boxes with a timebox in which
they occurred and a meaning in \tlaplus, solving for a logical total
order of actions. Unlike prior approaches based on linearizability 
checking,
\omnilink implements substep linearizability semantics, relies on a flexible
specification language, accelerates checking with user-customizable path pruning specifications,
and can detect violations due to weak memory effects between operations. 
%
%
%
Our evaluation validates WiredTiger, a state-of-the-art industrial
database storage layer; Balanced Augmented Tree (BAT), a
state-of-the art lock-free data structure from the research community;
and, ConcurrentQueue, a popular lock-free queue featuring aggressive
performance optimizations.
\omnilink detects injected and previously fixed bugs in the systems
under test.
It also uncovers a previously unknown bug in BAT that has been confirmed by its authors.



\end{abstract}




\maketitle







\section{Introduction}

Building correct concurrent systems remains a persistent software
engineering challenge. 
%
\emph{Trace validation} offers a promising middle ground between
lightweight testing and full verification. Rather than attempting to
prove correctness of all possible executions, trace validation asks
whether a \emph{concrete execution} of the system is consistent with
the behaviors allowed by a specification. A trace is collected from a
running system, and the validator checks if there exists some sequence
of specification actions that could have produced the same observable
outcomes. This style of validation has been explored in multiple
contexts~\cite{cirstea:validating,howard2024smartcasualverificationconfidential,multigrained,hackett_2025_tracelink}, from distributed protocols to storage systems,
because it combines two desirable properties: it scales to realistic
implementations, and it leverages the precision of formal models to
detect inconsistencies.

Despite its promise, 
most prior work relies on a \emph{single
serialized log}, which reduces concurrency to a sequential execution
and obscures the very interleavings most likely to trigger
bugs. The only trace validation approach that supports true concurrency is
TraceLink~\cite{hackett_2025_tracelink}, but it requires using the restrictive
PGo compiler~\cite{Hackett:2023}.

On the other hand, linearizability checking is an established field with a similar goal: checking whether a trace consisting of concurrent operations on a data structure can be interpreted as a valid sequence of atomic actions.
Conventional linearizability checking~\cite{winggong1993,lowe2016history} is done using specialized algorithms, and relies on checker-specific correctness specifications which are usually composed of known-correct primitives (registers, dictionaries, and similar objects).
For complex linearization semantics, there is no way to check that they themselves work as intended.

We present \textbf{\omnilink}, a new methodology that addresses the limitations of both trace validation and linearizability checking.
\omnilink checks specifications in \tlaplus, an established modelling language in which models can be automatically checked for self-consistency and arbitrary correctness properties.
This allows us to write more complex and robust linearizability semantics than currently possible.
\omnilink adopts the timeboxes and actions from linearizability checkers, and builds on the TLC model checker.
\omnilink also extends existing linearizability checking with the \emph{substep linearizability} semantics~\cite{hackett2026substep} to express non-linearizable behaviors, and a user-customizable path pruning performance optimization. \omnilink also introduces the \emph{burst mode} strategy to capture and check linearizability violations due to weak memory effects between operations.
We test \omnilink's performance characteristics and find that \omnilink scales better
than state-of-the-art linearizability checkers~\cite{porcupine, jia2023verilin}.


Given a \tlaplus specification that describes a concurrent system's API-level behavior and a timeboxed log that capture the concurrency of real executions, 
\omnilink provides linearizability solving via an instrumented TLC~\cite{tlatoolbox2019} model checker.
To explore rare thread interleavings, \omnilink integrates with Fray~\cite{li2025fray}, a
concurrency testing framework. To help with debugging, \omnilink supports
the \texttt{rr}~\cite{rr-project} record/replay tool.


We use \omnilink to validate three systems: the WiredTiger~\cite{wiredtiger} storage engine
underlying MongoDB, an advanced concurrent data structure called
Balanced Augmented Tree (BAT)~\cite{bat}, and the popular ConcurrentQueue~\cite{cameron314_concurrentqueue}, a non-linearizable lock-free queue with aggressive performance optimizations.
Our results show that \omnilink finds previously unknown bugs in BAT and WiredTiger, and it is able to reproduce known bugs.
\omnilink can also help develop new \tlaplus models using real system executions.

In summary, we make the following contributions:
\begin{itemize}[leftmargin=1em, itemsep=0.5ex, label=\(\star\)]

  

\item We describe \omnilink, a general-purpose trace validation
  approach. \omnilink adapts linearizability checking for the \tlaplus
  ecosystem, in order to validate \emph{unmodified} concurrent systems
  and support \emph{reuse} of existing \tlaplus models.

\item We propose the first algorithm for checking \emph{substep
  linearizability}~\cite{hackett2026substep}, which allows for finer-grained modelling of
  concurrent system behaviors.


\item We introduce \emph{path pruning}, a framework for users to
  express custom linearizability checking optimizations inexpressible
  using existing tools. We show that path pruning subsumes P-Compositionality~\cite{horn2015}.

\item We introduce the \emph{log vector optimization}, which improves
  the time and space complexity of linearizability checking
  implementations. This optimization makes validating longer traces
  more practical with \omnilink than with the existing state of the
  art~\cite{porcupine, horn2015, jia2023verilin}.

\item We propose \emph{burst mode}, the first runtime trace-based
  linearizability checking approach that detects violations due to
  weak memory effects between operations.


  



\end{itemize}

\section{Background}


\paragraph[TLA+ and System Modeling]{\tlaplus and System Modeling}
\tlaplus (Temporal Logic of Actions) is a formal specification
language for modelling and reasoning about concurrent
systems~\cite{SpecifyingSystems}. A \tlaplus specification describes a
system in terms of its \emph{state variables}, its \emph{initial state
predicate} (\texttt{Init}), and its \emph{next-state relation}
(\texttt{Next}), which defines how states evolve over time. The logic
supports reasoning about safety and liveness properties through
temporal operators.
%
%
\tlaplus specifications are written in a high-level, declarative style
that abstracts away details.

\paragraph{TLC Model Checking}
\emph{TLC} is an explicit-state model checker
that exhaustively explores the state space of a \tlaplus
specification~\cite{tlc1999}. TLC evaluates the specification's
possible behaviors to verify whether invariants and temporal
properties hold. Each explored state corresponds to a possible
configuration of the system's variables, and each transition
corresponds to an application of the \texttt{Next} relation.
%
TLC can detect violations such as deadlocks, invariant
breaches, or unexpected liveness failures. However, the state space
can grow exponentially with the number of variables and processes.

\paragraph{Linearizability}
Linearizability~\cite{herlihy1990linearizability} is a key correctness
condition for concurrent systems. 
It requires that every operation on a concurrent object
appears to take effect instantaneously at some point between its
invocation and response. This allows concurrent executions to be
understood as if they occurred sequentially and atomically.
%


\paragraph{Trace Validation}
Trace validation is the process of comparing observed system execution
traces against a specification to ensure conformance. For \tlaplus,
validation involves checking whether recorded traces from an
implementation can be interpreted as behaviors allowed by the
specification. 
%
Trace validation complements model checking: while TLC verifies
properties on the model, trace validation checks that the
\emph{implementation's behavior} adheres to the same logical
constraints. 

\paragraph{\omnilink's Assumptions}

\omnilink's Trusted Compute Base includes: the TLC model checker, its logging utility, \omnilink's \tlaplus linearizability logic, and auxiliary tooling that manages trace data.
Notably, \omnilink trusts neither the system being traced, the implementation logging statements, nor the model used for validation.
If the behavior of one of these is not consistent with the others, \omnilink will report a validation failure.
Section~\ref{sec:what-can-analyze} further details what \omnilink can and cannot analyze.

%


\section{Substep Linearizability}

In this section, we overview a generalization of linearizability called \emph{substep linearizability}~\cite{hackett2026substep}.
While linearizability has a lot of useful properties that make it intuitive to work with,
its requirement that operations behave atomically restricts its applicability.
For example, the semantics of concurrent objects such as exchangers~\cite{java_exchanger}, write-snapshot objects~\cite{cataneda2018intervallin}, ConcurrentQueue~\cite{cameron314_concurrentqueue}, and snapshot isolation database systems~\cite{snapshot_isolation95} cannot be defined using linearizability.
Substep linearizability~\cite{hackett2026substep} relaxes the atomicity requirement by allowing operations to take effect across multiple substeps, and can be applied to define an intuitive semantics for the non-linearizable objects listed above.

In Section~\ref{sec:substep-def}, we begin by presenting a simplified definition of substep linearizability on complete histories\footnote{In complete histories, all invoked operations have a matching response. These are the kinds of histories \omnilink checks.}. 
Then we give an example in Section~\ref{sec:non-lin-queue} of using substep linearizability to specify the behavior of the substep-linearizable ConcurrentQueue system.




%
%

\subsection{Substep Linearizability Definition}
\label{sec:substep-def}

We define a trace $T$ to be a set of timeboxes $\{T_1, T_2, …, T_N\}$. 
Each timebox $T_i$ has an action $a_i$, a start time $s_i$, and an end time $e_i$, such that $e_i > s_i$.
Each action consists of a unique identifier, an operation name, input arguments and a return value.
We define the \emph{real-time partial order} ($<_{RT}$) over the actions in the trace as follows: $a_i <_{RT} a_j$ if and only if $e_i < s_j$.

We can define linearizability using this formalism as follows.
First define a \emph{sequential trace} as a sequence of actions and a \emph{sequential specification} as a prefix closed set of sequential traces. 


\begin{definition}[Linearizability]
	A trace $T$ is \emph{linearizable} with respect to a sequential specification $S$ if and only if there is a linear extension of the real-time partial order that is in $S$.
\end{definition}

We extend this definition to substep linearizability by replacing `linear extension' with a new concept we call ``substep linearization''. A substep linearization is essentially a linear extension where actions can appear more than once and each action is tagged with a boolean indicating if the action has completed.

\begin{definition}[Substep Linearization]
A \emph{substep linearization} of $T$ is a sequence $L$ of $\langle Action, Boolean \rangle$ pairs such that:
\begin{enumerate}[leftmargin=*, noitemsep, topsep=0pt]
	\item $L$ contains the same set of actions as $T$,
	\item Each action in $T$ appears at least once in $L$ (but may appear multiple times)
	\item For the last occurrence of each action in $L$, the action’s boolean is set to true to indicate that it is done. For all other occurrences of this action, its boolean is set to false.
	\item The sequence $L$ preserves the real-time partial order in the following sense: if $a_i <_{RT} a_j$, then the last instance of $a_i$ appears before the first instance of $a_j$ in $L$.
\end{enumerate}
\end{definition}

\begin{definition}[Substep Linearizability]
  \label{defn:substep-lin}
	A trace $T$ is \emph{substep linearizable} with respect to a specification $S$ if and only if there is a substep linearization of $T$ that is accepted by $S$.
\end{definition}

Intuitively, substep linearizability allows actions to be broken up into multiple atomic substeps whose behavior is specified by $S$.
The specification $S$ can be defined by a state machine and a set of initial states. 
The edges of the state machine are labeled by $\langle Action, Boolean \rangle$ pairs. 
The state machine can be non-deterministic, meaning multiple outgoing edges from the same state can have the same label.
All states are accepting states. However, some transitions may not be possible at some states, which is why $S$ might reject some sequences.
A substep linearization $L$ is accepted by $S$ if and only if the state machine accepts $L$ starting from one of the initial states.
To simulate the definition of linearizability, we can require all edges in the state machine to have boolean value true. 
This disallows actions from appearing twice in $L$.
It is important for each action to have a unique id so that the state machine can distinguish between different operations that have the same name, arguments, and return value.
In Section ~\ref{sec:design}, we describe how we augment TLC to check this state machine.


\subsection{Example: A Substep Linearizable Queue}
\label{sec:non-lin-queue}

\begin{figure}[t]
\centering
\includegraphics[width=\columnwidth]{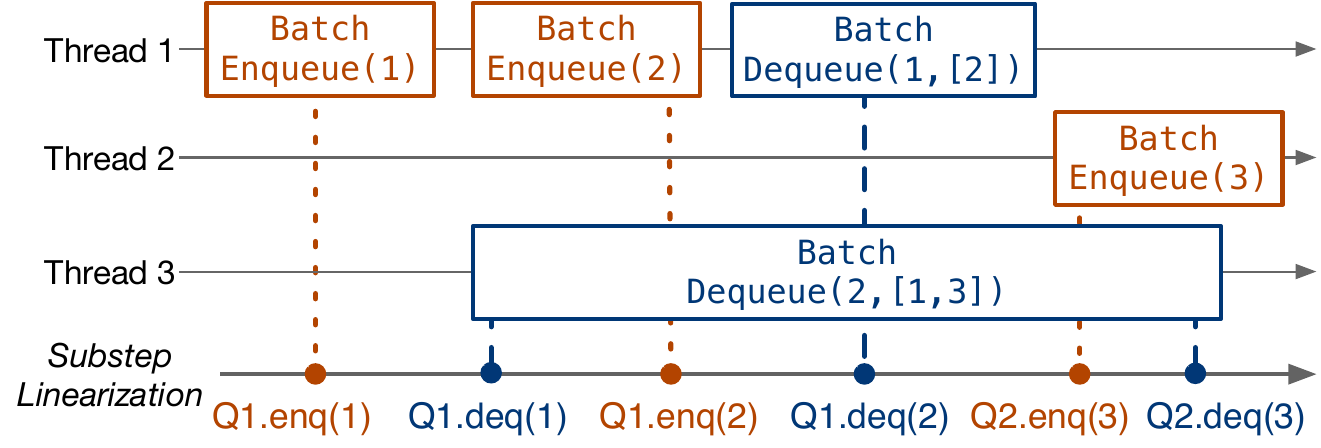}
\caption{Example three-threaded execution of a Desrochers's
  ConcurrentQueue. Batched operations are shown as timeboxes with time
  running from left to right.
  BatchEnqueue(i) enqueues element i. BatchDequeue(k,r) dequeues k
  elements and returns the list r of dequeued elements. Q.enq(i)
  enqueues element i into the underlying queue Q, and Q.deq(j) dequeues
  and returns j from the underlying queue Q.
  %
  The bottom timeline shows a \emph{substep linearization} for the
  sub-step decomposition of the batched operations above.}
\label{fig:timeboxes-non-lin}
\end{figure}

Achieving linearizability can be expensive, especially for batched operations.
For this reason, Desrochers's ConcurrentQueue~\cite{cameron314_concurrentqueue} supports \emph{non-linearizable} batched enqueue/dequeue operations (see example in Figure~\ref{fig:timeboxes-non-lin}).
To avoid contending on a single queue, ConcurrentQueue shards elements across $P$ subqueues, one per thread.
We now explain how to define the semantics of batched enqueue/dequeue operations on this data structure using substep linearizability.

A batch enqueue takes as input an ordered list of elements and the id of the current thread.
It has no return value.
Each substep of a batch enqueue takes the next element from its input list and enqueues it on the current thread's subqueue.
Once all elements from the input list have been enqueued, the batch enqueue is marked as completed.

%

A batch dequeue takes as input a positive integer $k$ indicating the number of elements to dequeue and returns a list of up to $k$ elements.
Each substep of a batch dequeue picks an arbitrary subqueue and either dequeues an element from it or observes that it is empty.
Once $k$ elements have been dequeued or all subqueues have been observed as empty, the batch dequeue is marked as completed and its return value is the list of elements dequeued over the course of its substeps (in the order they were dequeued).

The above behavior can be specified more formally as a non-deterministic state machine where each state tracks the contents of the $P$ subqueues as well as the progress of the ongoing operations. 
The non-determinism comes from the batch dequeue operation, since each dequeue substep is allowed to visit any subqueue.
The initial state is one in which all subqueues are empty and there are no ongoing operations.
In our experimental evaluation, we use \omnilink to validate that ConcurrentQueue is substep linearizable with respect to a weaker version of this specification, which allows dequeue operations to return fewer operations than requested without any constraints.

\section{\omnilink Design}
\label{sec:design}

This section describes \omnilink's architecture and formal semantics.
We build a decision procedure for substep-linearizability on top of the TLC model checker, which includes our novel log vector data structure.

\tlaplus views time as a strictly linear sequence of
logical states. A specification represents a set of action sequences,
where each \emph{action} operates atomically and transitions the
system from one state to another. Each action defines a precondition
on the current state and a postcondition describing all possible next
states. Importantly, \tlaplus provides no semantics for reasoning
about overlapping actions --- every behavior is a purely sequential
execution.

Prior efforts to compare implementation behavior with formal \tlaplus specifications are limited to either identifying precise times at which implementation events occurred~\cite{howard2024smartcasualverificationconfidential,cirstea:validating,multigrained}, or using heavyweight techniques like vector clocks~\cite{hackett_2025_tracelink}.
\omnilink bridges this gap by building on linearizability checking.
Linearizability checking for \tlaplus goes beyond conventional trace validation by inferring total event order, and \omnilink extends linearizability checking with substep linearizability and user-customizable dynamic path pruning.

\subsection{\omnilink architecture}

\begin{figure}[t]
\centering
\includegraphics[width=\columnwidth]{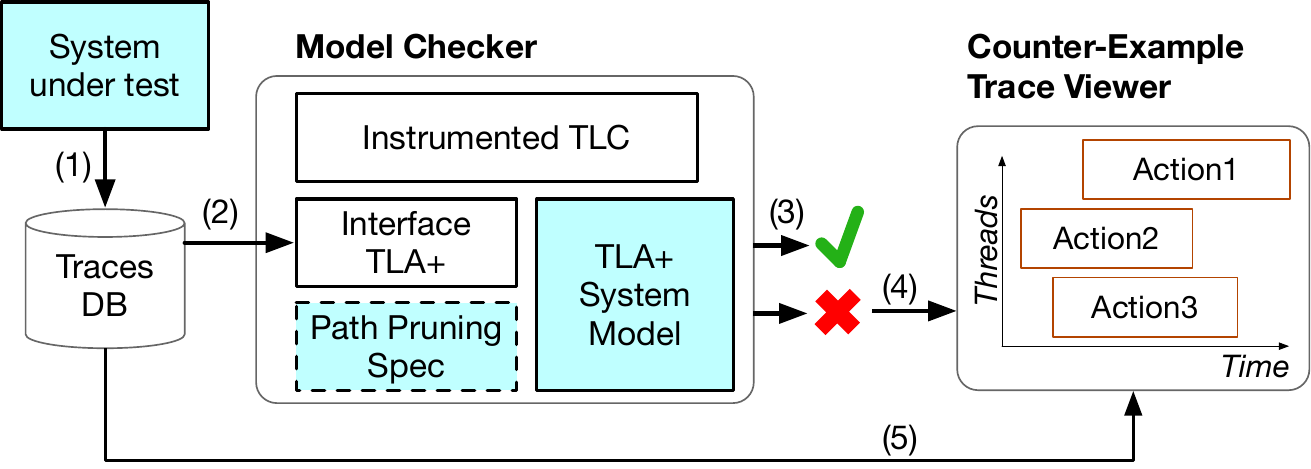}
\caption{\omnilink's overall design. The system's behavior is logged to a database.
  The logs are then read by a model checking process that either accepts or rejects them, based on both linearizability and conformance with a \tlaplus specification.
  The user can inspect the failed counter-examples using our custom viewer.
  Blue boxes are inputs provided by the user to \omnilink. The non-blue pieces are provided by \omnilink.}
\label{fig:arch}
\end{figure}

\omnilink's architecture follows two stages: (1) execute and log implementation operations, and then (2) analyze the recorded traces to determine if they match a given \tlaplus model.
Figure~\ref{fig:arch} breaks these steps down further, with blue boxes representing inputs provided by the user.

On the implementation side, \omnilink requires logs describing the application component of interest.
For each event, the information required is start time, end time, \tlaplus action name, and any parameters needed by the \tlaplus action.
This information unambiguously describes which part of the system under test is supposed to implement which part of the \tlaplus model.
It is the user's responsibility to build a mapping from implementation code to \tlaplus model actions.
A common pattern is to write a wrapper program that randomly calls API functions on the system under tests, but \omnilink does not assume this.
The only requirement is to self-consistently log the information needed.
We provide both a C library and an experimental Java library to assist in emitting \omnilink-compatible logs.

Given an instrumented program, \omnilink ingests the log stream into its database.
Given a complete log stream, \omnilink uses an instrumented copy of the TLC model checker to perform a reachability check: there must be a path of valid \tlaplus model actions, starting at the model's initial state(s), covering all recorded log entries, and respecting timebox order.
If such a path exists, \omnilink will accept the log as valid.
If no such path exists, the log is rejected with the covered state space as a counterexample.
Since \omnilink is performing a dynamic analysis, the validity of one log does not imply the validity of any other logs.
For triage purposes, \omnilink provides a viewer (TraceView) that allows interactive exploration of counterexamples, showing the timebox sequences and the logged events in the context of the TLC state graph. 

\subsection{Data Model}

\omnilink must provide a data model that is both amenable to representing implementation logs, and to interacting with \tlaplus.
Our design bridges this gap by observing that the existing explicit-state \tlaplus model checker, TLC, uses a data model that is already similar to schemaless data storage formats, such as JSON.

\omnilink's value model is a JSON-like system with booleans, integers and strings as primitives.
\omnilink can also represent sets, tuples (lists), and functions (dictionaries).
As a result, \omnilink's log entries can be ingested 1-1 into the \tlaplus ecosystem, and the format is similar enough to popular formats that it can easily be logged from any implementation.

\begin{figure}
\small
\begin{bnf}
\textit{log} ::= \textit{action}$^*$
;;
\textit{action} ::= (\textit{op}(\textit{value}$^*$), \textit{start-time}, \textit{end-time})
;;
\textit{op} ::= \underline{name}
;;
\textit{value} ::=
| \underline{boolean}
| \underline{integer}
| \underline{string}
| $\left\{\textit{value}^*\right\}$
| $\left<\textit{value}^*\right>$
| $\left\{(\textit{value} \to \textit{value})*\right\}$
;;
\textit{start-time} ::= \underline{time}
;;
\textit{end-time} ::= \underline{time}
\end{bnf}
\caption{Grammar of traces used by \omnilink.}
\label{fig:trace-grammar}
\end{figure}

Figure~\ref{fig:trace-grammar} shows \omnilink's model for traces and
values.  The root is a log, which contains a collection of actions.
Actions name a \tlaplus action, list its parameters, and give a start
and end time that define the timebox for linearization.
\omnilink's value model is a simple JSON-like system with booleans,
integers and strings as primitives.  Additionally, \omnilink can
represent sets, tuples (lists), and functions (dictionaries).

\subsection{Model Checking Extensions}

\omnilink extends \tlaplus model checking.
In \tlaplus, any given state is modeled as a function $s: \textit{Variable} \to \textit{Value}$ which assigns every state variable in the model to some representable value.
The conventional \tlaplus state graph is the binary relation $s\ \mathcal{T}\ s'$, for any states $s$ and $s'$.
An explicit state model checker would generate a set of initial states, and then explore paths $s\ \mathcal{T}\ s'\ \mathcal{T}\ s'' \dots$ in order to disprove \tlaplus properties.

\omnilink makes three changes to this process: (1) states are tagged with program order information; (2) the transition relation considers logged actions and linearizability restrictions; and (3) state space traversal checks reachability of the end of the log, via some path that includes every log entry.

\subsubsection{States with Vectors}

An \omnilink state can be constructed from a regular \tlaplus state $s$ and a log vector $v \in \mathcal{N} \times \mathcal{N} \times \mathcal{N} \dots$, forming the tuple $\left<s, v\right>$.
The size of $v$ is constant for a given log, and is derived from the number of unique threads/processes recorded in the log.
Each thread has a unique number\footnote{In practice, we directly use the OS thread/process ID for this purpose.} $i$, for which $v(i)$ indexes program order within that thread.
For instance, given a log with 3 threads and an initial vector $v = \left<0, 0, 0\right>$, thread 2 completing its first action would produce $\texttt{inc}(v, 2) = \left<0, 1, 0\right>$.
Thread 3 completing its first action after thread 2 would produce $\texttt{inc}(\texttt{inc}(v, 2), 3) = \left<0, 1, 1\right>$. 
Generally, we use $\texttt{inc}(v, i)$ to indicate a vector ``advancing at $i$''.

Since $v$ identifies a unique coordinate within the log, we use them to refer to information in the log. 
For instance, $\texttt{action}(v, i)$ is the action logged by thread $i$ at step $v(i)$.

When studying related work, we found that state of the art linearizability checker designs~\cite{horn2015,porcupine} use a different structure to track accepted log events.
They use a bit vector of size equal to the log being validated and maintain copies of it per log event, leading to quadratic space complexity, and linear overhead in constructing each copy.
\omnilink's log vector scales in number of threads, not log size, which is an algorithmic advantage for logs with a high event to thread ratio.

\subsubsection{Augmented Transition Relation}

\omnilink's transition extends the conventional \tlaplus relation $\mathcal{T}$.
Before considering linearizability, $\mathcal{T}$ must first be extended to account for action choice and sub-steps.

While an unconstrained next-state relation may choose any action in the \tlaplus model at any time, \omnilink needs to identify whether a specific named action is possible in a given state.
So, we extend the transition relation as $s\ \mathcal{T}_{\texttt{op(...)}}\ s'$, which is true iff $s\ \mathcal{T}\ s'$, and $s'$ was reached by performing specifically $\texttt{op(...)}$.
In \tlaplus, we express this by directly evaluating $\texttt{op(...)}$ as the next-state definition, rather than the usual $\texttt{Next}$ definition of all possible actions.

\begin{listing}
\begin{pluscal}
op(...) ==
  \/ op_A(...)
  \/ op_B(...)
  \/ op_Finished(...)
\end{pluscal}
\caption{Illustration of \tlaplus sub-step specification format.}
\label{lst:op-ab}
\end{listing}

To support sub-steps, we must be able to identify whether the action's sub-step boolean from Definition \ref{defn:substep-lin} is true or false.
We do this by allowing $\texttt{op(...)}$ to be defined as a disjunction of sub-actions.
Any one of those actions may be an interpretation of $\texttt{op(...)}$, but only the last one, \\$\texttt{op\_Finished(...)}$ in Listing \ref{lst:op-ab}, is the final sub-step.

By convention, we name substeps by adding suffixes to the base action name, in this case $\texttt{op}$.
Listing \ref{lst:op-ab} shows the \tlaplus definition of two non-final sub-steps $\texttt{A}$ and $\texttt{B}$, and the final sub-step $\texttt{Finished}$.
All definitions take the same parameters, and collectively define the modeled behavior of the single log entry $\texttt{op}(...)$.

Based on this definition, we further instrument our transition function to classify whether the action's last substep was taken.
We write $s\ \mathcal{T}_{\texttt{op(...)}} \langle s', b \rangle$, where $b$ is the boolean from Definition \ref{defn:substep-lin}.
For Listing \ref{lst:op-ab}, $b$ will be true if $\texttt{op\_Finished(...)}$ produced $s'$, and false if $\texttt{op\_A}(...)$ or $\texttt{op\_B}(...)$ produced $s'$.

For convenience, we also write $\langle s, v \rangle\ \mathcal{T}_\texttt{op(...)} \langle s', v' \rangle$, such that given $s\ \mathcal{T}_{\texttt{op(...)}} \langle s', b \rangle$, for some $i$ such that $\mathtt{action(v, i) = \mathtt{op(...)}}$, if $b$ is true then $v' = \mathtt{inc}(v, i)$ and if $b$ is false then $v' = v$.

\subsubsection{Linearizable Reachability Algorithm}

\begin{listing}
\begin{align*}
& {\scriptstyle 1\ \ } \texttt{stack} \leftarrow \texttt{push\_initial\_states}() \\
& {\scriptstyle 2\ \ }\texttt{visited\_states} \leftarrow \texttt{stack}.\texttt{all\_elements}() \\
& {\scriptstyle 3\ \ }\textbf{while}\ \texttt{stack}.\texttt{non\_empty}() \\
& {\scriptstyle 4\ \ }\quad \left<s, v\right> \leftarrow \texttt{stack}.\texttt{pop}() \\
& {\scriptstyle 5\ \ }\quad \textbf{forall}\ \langle s', v' \rangle \in D(S(s, v))\quad \\
& {\scriptstyle 6\ \ }\quad \quad \textbf{if}\ \left<s', v'\right> \notin \texttt{visited\_states} \\
& {\scriptstyle 7\ \ }\quad \quad \quad \texttt{stack}.\texttt{push}(\left<s', v'\right>) \\
& {\scriptstyle 8\ \ }\quad \quad \quad \texttt{visited\_states} \leftarrow \texttt{visited\_states} \cup \{\left<s', v'\right>\} \\
& {\scriptstyle 9\ \ }\quad \quad \textbf{end if} \\
& {\scriptstyle 10\ \ }\textbf{end while}
\end{align*}
\caption{\omnilink's linearizable reachability algorithm.}
\label{lst:reachability-alg}
\end{listing}

\omnilink's reachability checking algorithm incorporates vector-marked states and the augmented state transition relation into an otherwise standard depth-first search, as shown in Listing \ref{lst:reachability-alg}.
The \tlaplus model's initial states are pushed onto the stack on line 1, marked with $v = \left<0, 0, \dots\right>$.
Then, the loop starting on line 3 examines states in stack order.
The conditional on lines 6--9 pushes only previously unseen states onto the stack.

The core transition function is referenced on line 5.
Given some existing state $s$ and vector $v$, it produces a next-state $s'$ and next-vector $v'$.
$D$ is the path pruning restriction discussed in Section~\ref{scn:dpor-formal}; for now assume $D(S(s, v)) = S(s, v)$.
The precise conditions for evaluating $S(s, v)$ are:
\begin{align*}
  \langle s', v' \rangle \in S(s, v) \text{ iff } \exists i : & \; (\nexists j : \texttt{action}(v, j) <_{RT} \texttt{action}(v, i)) \\
    & \land \langle s, v \rangle\ \mathcal{T}_{\mathtt{action}(v, i)} \langle s', v' \rangle
\end{align*}

To discover new states, $S(s, v)$ chooses any thread $i$ for which $\texttt{action}(v, i)$ does not contradict real time and is possible in $s$ according to $\mathcal{T}$.
The real time condition means there must not exist any other thread id $j$ whose operation at $v_j$ is real-time ordered before the operation $v_i$.
If there existed such an operation, it should be sequenced before $v_i$.
By eliminating all such non-linearizable choices, \omnilink is only able to explore paths that are linearizable.
Note that this condition is equivalent to requiring all $v_j$ to either happen-after or be concurrent with $v_i$.
Otherwise, $s'$ and $v'$ are constrained to be any valid instance of $\mathcal{T}$.
In practice, $S(s, v)$ is implemented by using the TLC model checker's next-state function as a subroutine given real-time valid choices of $i$ and post-processing the list of candidate next-states.

For convenience, we write the set of all possible executions of operation op in state $s$ with vector $v$ as $S_\text{op}(s, v) = S(s, v) \cap \{ \langle s', v' \rangle \text{ where } \langle s, v \rangle\ \mathcal{T}_\text{op} \langle s', v' \rangle \}$.

\subsection{Path Pruning}
\label{scn:dpor-formal}

When performing linearizability checking, it is desirable to reduce the range of considered behaviors as far as possible.
If we encounter a state with two transition, A and B, and the user can prove that taking transition A leads to a linearization only if taking transition B leads to a linearization.
Then, we do not need to explore transition A and can prune it out of the state graph. 

\omnilink offers \emph{path pruning} to support this.
Given any state and its set of transitions (each labeled by a candidate operation), the user can identify a subset of transitions to prune away.
The simplest example is read-only operations that have no effect on system state.
Sequencing those operations before other candidates is always safe, since it results in an unchanged state where every operation that was originally possible remains so.
Operations that fulfill this condition while having side-effects must only have local side-effects, such as adjusting thread or process-local data.

Path pruning is related to both P-Compositionality~\cite{horn2015} and Dynamic Partial Order Reduction~\cite{dpor-2025} (DPOR).
Path pruning can be used to implement P-Compositionality, but it is less powerful than a general-purpose DPOR algorithm, which is capable of speculation and/or lookahead in the state graph.
Path pruning is a novel addition to both \tlaplus and general-purpose linearizability checking tools.

\subsubsection{Formal Definition}

Given $S(s, v)$, we require some user definition $D(S(s, v)) \subseteq S(s, v)$.
The goal of $D$ is to soundly reduce $\left|S(s, v)\right|$, which it may do by removing redundant elements.
We use $D$ on line 5 of Listing \ref{lst:reachability-alg}.
That is, for any transition enumerated by \omnilink, only transitions not discarded by $D$ will be pushed onto the stack, and therefore participate in the reachability check.
$D$ may trivially be the identity function, in which case \omnilink does not perform path pruning.

To evaluate soundness of a given specification of $D$, we reason about $\mathcal{L}$, the set of linearizable paths for any given log.
Each path $L \in \mathcal{L}$ has the form $v_0, ..., v_n$, where each element is a log vector.
For every $L$ there exists at least one corresponding state sequence $s_0, ..., s_n$, such that for all $k \in 1..n$, $\langle s_k, v_k \rangle \in S(s_{k-1}, v_{k-1})$.
$v_0$ is the initial vector $\langle 0, 0, ... \rangle$, and $s_0$ is a valid initial state defined in the \tlaplus model.

In this context, Definition~\ref{defn:sound-pruning} gives the condition for soundly pruning paths.
Intuitively, a choice of $D$ must never shorten the longest possible linearization of a given log.
This ensures that linearizable logs remain linearizable after pruning and that \omnilink returns one of the longest possible linearization for non-linearizable logs.


\begin{definition}[Sound Path Pruning]\label{defn:sound-pruning}
	Let $n$ be the length of the longest possible linearization in $\mathcal{L}$. There exists a linearization of length $n$  in $\mathcal{L}$ that does not use any of the pruned transitions. More precisely, there exists $v_1, ..., v_{n} \in \mathcal{L}$ and corresponding states $s_1, ..., s_{n}$ such that $\langle s_{i+1}, v_{i+1}\rangle \in  D(S(s_i, v_i))$ for all $i$.
\end{definition}

\subsubsection{Identity Pruning}

To demonstrate a pruning rule that is sound, consider this definition of $D_\text{ident}$:
\begin{align*}
  D_\text{ident}(S(s, v)) & =  S_{\mathtt{action}(v, i)}(s, v) & \text{ if } \exists i : \forall \langle s', v' \rangle \in S_{\mathtt{action}(v, i)}(s, v) : s = s' \\
  D_\text{ident}(S(s, v)) & = S(s, v) & \text{ otherwise}
\end{align*}

The purpose of $D_\text{ident}$ is to perform operations that have no effect (i.e. read-only operations) as soon as they become valid transitions.
If one or more operations have no effect, constrains the set of next-states to include arbitrarily any such operation, linearizing all non-effectful operations as early as possible.
Otherwise, the naive algorithm would consider all interleavings of no-effect operations despite them being functionally identical.
The side-condition requires that all of a chosen thread's possible next steps must have no effect, which is important for soundness.
Consider when a thread $i$ has two possible behaviors and one has no effect, pruning the effectful step may be unsound.
Operations by the same thread are inherently mutually exclusive, so choosing only the non-effectful case will prevent the effectful case from being considered.

\begin{theorem}\label{thm:ident-sound}
  $D_\text{ident}$ is a sound path pruning.
\end{theorem}

\begin{proof}
  Let $v_1, ..., v_n \in \mathcal{L}$ be one of the longest linearizations in $\mathcal{L}$, and let $s_1, ... s_n$ be its corresponding sequence of states.
  If this linearization uses no pruned transitions, then $D_\text{ident}$ is sound.
  Otherwise, we show how to remove pruned transitions one by one until we get a linearization of length $n$  that uses no pruned transitions.
  Let $i$ be the index of a first pruned transition used by $v_1, ..., v_n$. That is, $\langle s_{i+1}, v_{i+1} \rangle \in S(s_i, v_i) \setminus D_{ident}(S(s_i, v_i))$ and for all $j < i$, $\langle s_{j+1}, v_{j+1} \rangle \in D_{ident}(S(s_j, v_j))$.
  We will construct a linearization $v'_1, ..., v'_{n}$ where the prefix $v'_1, ..., v'_{i+1}$ does not use any pruned transitions. Applying this transformation up to $n$ times allows us to remove all pruned transitions without changing the length of the linearization. The transformation is as follows.

  Since some transition from state $\langle s_i, v_i \rangle$ was pruned away, by definition of $D_{ident}$, this means  $D_\text{ident}(S(s_i, v_i)) =  S_{\mathtt{action}(v_i, a)}(s_i, v_i)$ for some process id $a$ such that $\forall \langle s, v \rangle \in S_{action(v_i,a)}(s_i,v_i): s = s_i$. In other words, the operation $op$ that process $a$ wishes to perform next according to the log vector $v_i$ is a read-only operation and $op$ is a valid operation to perform on state $s_i$ according to the TLA+ spec, and $D_\text{ident}$ pruned away all transitions other than those labeled by $op$.
  Suppose for contradiction that the linearization $v_1, ..., v_n$ does not follow any transition labeled by $op$. Then process $a$ does not perform any operations after $v_i$ in this linearization and we can construct a longer linearization $v_1, ..., v_i, w_i, w_{i+1} ..., w_{n}$ with corresponding states $s_1, ..., s_i, s_i, ...,s_n$ by performing $op$ between $v_i$ and $w_i$ and defining the $w$'s as follows. For all $j \in [i, n]$, $w_j$ equals $v_j$ except with the log vector entry for process $a$ incremented by one.
  This is a valid linearization since applying $op$ at state $\langle s_i,  v_i \rangle$ results in state $\langle s_i,  w_i \rangle$ and for all $j \in [i+1, n]$, if the transition from $\langle s_{j-1}, v_{j-1} \rangle$ to $\langle s_{j}, v_{j} \rangle$ is allowed by $S$, then the transition from $\langle s_{j-1}, w_{j-1} \rangle$ to $\langle s_{j}, w_{j} \rangle$ is also allowed by S because none of these transitions are labeled by an operation by process $a$, and incrementing the log vector entry for $a$ can only increase the possible transitions by other processes.
  Therefore, we have constructed a linearization of length larger than $n$ which contradicts our choice of $n$.

  So, the linearization $v_1, ..., v_n$ follows a transition labeled by $op$ and it must do so between $v_k$ and $v_{k+1}$ for some $k > i$. We can construct a new linearization $v_1, ..., v_i, w_{i+1}, ..., w_{k},v_{k+1},...,v_{n} \in \mathcal{L}$ where $op$ is applied between $v_i$ and $w_{i+1}$ so that this transition uses an edge that is kept by $D_{ident}$.
  For $j \in [i+1, k]$ we define $w_j$ to be $v_{j-1}$ with the log vector entry for process $a$ incremented by one. The states corresponding to this new linearization are $s_1, ..., s_i, s_i, ..., s_{k-1},s_{k+1},...,s_n$.
  All that remains is to show that $w_{i+1}, ..., w_{k},v_{k+1},...,v_{n}$ and $s_i, ..., s_{k-1},s_{k+1},...,s_n$ are a valid sequence of states according to $S$ starting from state $\langle s_i. v_i \rangle$. The immediate next state $\langle s_i, w_{i+1} \rangle$ is valid because it is in $D_\text{ident}(S(s_i, v_i))$. For all $j \in [i+2, k]$, if the transition from $\langle s_{j-2}, v_{j-2} \rangle$ to $\langle s_{j-1}, v_{j-1} \rangle$ is allowed by $S$, then the transition from $\langle s_{j-2}, w_{j-1} \rangle$ to $\langle s_{j-1}, w_{j} \rangle$ is also allowed by S because none of these transitions are labeled by an operation by process $a$ and incrementing the log vector entry for $a$ can only increase the possible transitions by other processes. The transition from $\langle s_{k-1}, w_k \rangle$ to  $\langle s_{k+1}, v_{k+1} \rangle$ is valid because $\langle s_{k+1}, v_{k+1} \rangle = \langle s_{k}, v_{k+1} \rangle \in S(s_{k-1}, v_{k-1}) \subset S(s_{k-1}, w_k)$. Thus proving that $v_1, ..., v_i, w_{i+1}, ..., w_{k},v_{k+1},...,v_{n}$ is a valid linearization of length $n$ where the first $i$ transitions are kept by $D_{ident}$, as required.
 
%
%
\end{proof}

\subsubsection{Generalized Identity Pruning}\label{scn:nb-pruning}

It is possible to generalize the idea above so that it applies to more than just read-only operations. The core argument in the proof of Theorem~\ref{thm:ident-sound} is the following: if the sequence of operations $op_1, ..., op_{i-1}, op_i, op_{i+1} ..., op_{j-1}, op_j, op_{j+1}, ..., op_n$ is a linearization and $op_j$ is a read-only operation that is ready to execute after $op_1, ..., op_{i-1}$ (according to both the TLA+ spec and real-time ordering), then $op_j$ can be moved immediately before $op_i$ and the resulting sequence $op_1, ..., op_{i-1}, op_j, op_i, op_{i+1} ..., op_{j-1}, op_{j+1}, ..., op_n$ would still be a valid linearization (which respects real-time order and the TLA+ specification). This means read-only transitions can be taken as soon as they are available when exploring the state space.

For this argument to work, $op_j$ need not be read-only --- it may modify the TLA+ state as long as the following \emph{generalized identity} properties holds for $op_j$: (1) for any sequence of operations $op'_1, ...., op'_{j-1}, op_j, op'_{j+1}, ..., op'_k$ that is accepted by the TLA+ spec, we can remove $op_j$ and the resulting sequence would still be accepted by the TLA+ spec, (2) conversely, if the sequence $op'_1, ...., op'_{j-1}, op'_{j+1}, ..., op'_k$ and $op'_1, ...., op'_{j-1}, op_j$ are both accepted by the TLA+ spec, then $op'_1, ...., op'_{j-1}, op_j, op'_{j+1}, ..., op'_k$ is accepted by the TLA+ spec.
This allows us to remove $op_j$ from $op_1, ..., op_n$ and add it before $op_i$ in the core argument above without violating the TLA+ spec. The argument for why this change does not violate real-time ordering remains the same.
We make use of this generalization in the evaluation section.

\subsubsection{Relation to P-Compositionality}\label{scn:relation-to-p-comp}

In concrete linearizability checking, P-Compositionality~\cite{horn2015} is the state of the art in state space explosion optimizations.
It relies on the global logical independence between disjoint subsets of operations in a log, partitioning those subsets into independent linearizability checking problems.
Any P-Compositionality optimization can be expressed as a path pruning optimization, but not the other way around.

For instance, P-Compositionality cannot express the identity optimization from Theorem~\ref{thm:ident-sound}.
It is possible to approximate such an optimization if a specific log entry type can be identified as never having an effect.
In that case, it is possible to partition no-effect operations into their own group.
However, identity pruning applies to operations that \emph{may have had an effect but emergently didn't}.
This includes cases where in different possible linearizations of the same log, one linearization gives an operation an effect, and the other one doesn't.
Such cases cannot be resolved by any possible static log partition.
However, path pruning can express any instance of P-Compositionality.
We show a proof of this in Section~\ref{sec:p-comp-path-pruning}.

A key limitation of path pruning is its inability to express optimizations over data other than $S(s, v)$.
This precludes taking advantage of operations being commutative, for example.
We believe these limitations could be addressed by adding support for lookahead and/or backtracking to our path pruning design.
This idea is inspired by how dynamic partial order reduction~\cite{dpor-2025} is implemented in model checkers.
We leave this improvement as future work.

\subsubsection{P-Compositionality as Path Pruning}\label{sec:p-comp-path-pruning}

Consider a partition of operations $P_1,...,P_n$ given from a P-Compositionality optimization.
By definition of P-Compositionality, we have that the log is linearizable if and only if each partition is.
For a given set $S(s,v)$, let $S_k(s, v) = \bigcup_{o \in P_k} S_o(s, v)$, in other words the set of transitions in $S(s,v)$ using operations in $P_k$.
We define $K(v,i)$ as the unique $k$ such that the operation performing $\mathtt{action}(v,i)$ is in the partition $P_k$.
Finally, we define $Min(s,v)$ as the $i$ such that the response of the operation performing $\mathtt{action}(v,i)$ is minimum among all valid operations at $S(s,v)$.
Now we give a path pruning optimization equivalent to the P-Compositionality optimization on the partition $P_1,...,P_n$.
\begin{align*}
  D_\text{part}(S(s, v)) & = S_k(s, v) & \text{ where } k = K(v,Min(s,v))
\end{align*}

That is, $D_\text{part}$ is the set of transitions whose operation is within the partition that contains the operation with the current minimum response.

\begin{theorem}\label{thm:part-sound}
  $D_\text{part}$ is a sound path pruning
\end{theorem}

\begin{proof}
  We show there exists a longest sequence $L = v_0, ..., v_p \in \mathcal{L}$ such that, for $i \in [0,p)$, $\langle s_{i+1}, v_{i+1} \rangle \in D_\text{part}(S(s_i, v_i))$.

  Take some arbitrary linearization $L'\in\mathcal{L}$. By P-Compositionality, we know that $L'$ can be split into $n$ independent linearizations $L_1, ..., L_n$ such that each $L_k$ is linearizable on its own.
  We must prove that $D_\text{part}$ traverses at least one order-preserving interleaving $L$ of all $L_k$ sub-sequences.

  Since $D_\text{part}(s, v) = S_k(s, v)$ where $k = K(v, Min(s, v))$, we must show that $S_k(s, v)$ always explores an element of $L_k$.
  That is, it doesn't matter which partition is explored at each step since they are all independent, but each step must ensure it chooses at least the next element of $L_k$.

  Consider some $s$ and $v$ without loss of generality.
  For contradiction, assume there exists some operation $p$ whose action is $o \in P_k$, which is possible in $s$ and whose timebox is concurrent with all steps in $S_k(s, v)$, but which is not reflected in $S_k(s, v)$.
  This means there exists a partition $k'$, such that an operation $p'$ possible in $S_{k'}(s, v)$ has a timebox that happens-before $p$.
  For $p'$ to happen-before $p$, its timebox must end before the timebox of operation $p$ starts, contradicting the premise that $k = K(v,Min(s,v))$.

  Since $S_k(s, v)$ will always explore all possible steps in partition $k$, one of those steps must be the next element of $L_k$.
  By repeating this process, the maximal linearization $L$ must eventually be constructed.
\end{proof}

In summary, we described path pruning, a novel user-customizable
performance optimization. In Section~\ref{sec:eval-path-pruning} we
evaluate the impact of path pruning.

\subsection{Weak Memory and Burst Mode}\label{sec:burst-mode-design}

Weak memory semantics are an important aspect of building and understanding concurrent systems.
There is an inverse relationship between potential performance and the synchronization of memory accesses, so modern hardware and compilers give developers access to increasingly weak memory semantics~\cite{sarkar2011understanding,sewell2010x86}.
At the same time, these weak memory semantics allow subtle, hard to reason about behaviors.
Many data structures take advantage of weak memory semantics internally while providing an interface that is linearizable, or at least substep-linearizable.
This makes it desirable for \omnilink to be able to check traces from such systems, and motivates the burst mode feature.

The key problem burst mode solves is measuring the weak memory effects to begin with.
Weak memory semantics allow reordering memory reads and writes up to specific constraints, or barriers.
Measuring a timestamp induces such a barrier, so reading a timestamp in between API calls, as is the expected practice when using any linearizability checker, inherently disables weak memory reorderings between operations that would otherwise be present.
Our solution is to skip measuring some timestamps, and make the analysis tool robust to this omission.
In essence, burst mode allows a ``burst'' of sequential events by the same thread to share the same timebox, meaning we only need the first event's start time and the last event's end time, and can freely execute uninterrupted back to back API calls.
Note that we still maintain the relative ordering of operations within a burst, despite them having the same timebox.

Consider the the definition $S(s, v)$, as used by the algorithm in Listing~\ref{lst:reachability-alg}.
To make progress in the log, it requires that there is no timebox referenced by the log vector that happens-before a chosen step's timebox.
If the same thread performs a sequence of actions with the same timebox, this is still true: each local step is followed by another step with the same timebox, which updates the system state but leaves the thread's timebox unchanged.
Steps concurrent with one element of a burst are therefore concurrent with all of them, and \omnilink is free to infer when each step took place in the same way it reasons about substeps.

While the log is less precise with missing timestamps, the ability to measure weak memory effects at all is useful.
Our experiment in Section~\ref{sec:eval-weak-mem-bug} gives a practical example, where we detect undesirable interprocedural weak memory effects using this technique.

\section{What \omnilink Can and Cannot Analyze}
\label{sec:what-can-analyze}

Here we detail what \omnilink can and cannot analyze.

\subsection{What \omnilink Can Analyze}

\paragraph{Non-Linearizable Systems}
\omnilink uses substep linearizability as its specification semantics, which is state of the art in sequentially modelling non-linearizable systems.

\paragraph{Missing Instrumentation}
If a system's behavior is only partially instrumented, \omnilink can flag a validation failure showing unexplained changes due to the uninstrumented parts.
This makes it possible to diagnose missing instrumentation.
In this sense, \omnilink does not rely on the system's instrumentation.

\paragraph{Concurrency Control Bugs}
Though the details of low-level concurrency control are not usually relevant to an abstract model, the consequences of not implementing concurrency control correctly will manifest as validation failures.
This can be true especially in the context of complex, subtle implementation optimization techniques.
If two concurrent operations have no valid total order when cross-referenced against each other, then there might be a concurrency control bug in the implementation.

\paragraph{Data Integrity Bugs}
Since \omnilink usage involves consistently recording the state of system components, any state transitions that do not match specification behavior will not validate.
These mismatches can expose both model and implementation inaccuracies.

\paragraph{Weak Memory Models}
\omnilink's burst mode will record weak memory behavior.
Advanced data structures that use weak memory semantics internally but have a substep-linearizable API are suitable for checking with \omnilink.

\subsection{What \omnilink Cannot Analyze}

\paragraph{Distributed System Behaviors}
While it is not theoretically impossible to validate distributed system behaviors with \omnilink, we rely on precise clock measurements to constraint linearization.
We leave evaluation of \omnilink on distributed systems as future work.

\paragraph{Non-Modeled Behavior}
A model that is not precise enough will cause \omnilink to miss bugs, since imprecision allows model checking to accept a wide range of traces, including buggy ones.
A consequence of this is that precise and wrong models are more valuable in practice than more permissive models.
When a restrictive model fails validation, it allows the developer to triage and understand the issue, whereas a permissive model that silently accepts an unusual behavior provides no insight.

\paragraph{Non-Traced Behavior}
\omnilink cannot detect bugs about which it has no information.
As a result, system components whose behavior does not reach \omnilink cannot be evaluated for correctness.
It is possible to indirectly validate something whose behavior is partially reflected in the traces, however.

\paragraph{Coincidentally Hidden Behavior}
Like any tracing based method, it is possible for a problematic behavior to occur, but for evidence of that behavior to be hidden.
For instance, a system with an incorrect synchronization mechanism can coincidentally perform actions that pass validation.
The mitigation is to capture multiple traces. We can induce more diverse traces by fuzzing inputs to the system and thread interleavings.

\section{\omnilink Implementation}
\label{sec:impl}

\omnilink is implemented as a Scala library, Mill build tool plugin, and set of PostgreSQL stored procedures.
The instrumented copy of the TLC model checker is packaged separately as a Scala library on which \omnilink depends.
In total, \omnilink is 7571 lines of Scala, 362 lines of SQL, 221 lines of Nix, and 90 lines of Java.
The logging support library is 1933 lines of C.

For the systems we evaluate, table \ref{tbl:systems} shows the per-system \tlaplus and C++ code.

\omnilink includes a TraceView tool to support understanding of
concurrent traces during debugging of \omnilink violations. TraceView
is described in Appendix~\ref{sec:understanding}. The TraceView
subproject is 1177 lines of Scala.

\section{Evaluation}
\label{sec:eval}

We evaluate \omnilink by asking the following research questions:
\begin{itemize}[leftmargin=*, noitemsep, topsep=0pt]
  \item \textbf{RQ1}: Does \omnilink's design help it find more bugs compared to related work?
  \item \textbf{RQ2}: How easy to use is \omnilink compared to other tools in terms of expressivity, implementation code, and specification effort?
  \item \textbf{RQ3}: Does the log vector optimization improve linearizability checking performance over the state of the art?
  \item \textbf{RQ4}: Does path pruning improve linearizability checking performance?
  \item \textbf{RQ5}: Can burst mode expose weak memory bugs in practice?
\end{itemize}

To answer these questions, we validated 4 different systems using
\omnilink.  We describe those systems, the validation workflow, the
bugs \omnilink could detect, and \omnilink's performance characteristics
while doing so. We also describe out experience in integrating
\omnilink with the Fray~\cite{li2025fray} concurrency testing platform, and 
with the rr~\cite{rr-project} record/replay framework.



\begin{table}
\centering
\small
\caption{Description of systems that we used to evaluate \omnilink.
Includes system implementation LOC, fuzzer LOC (Fuzz), specification \tlaplus, and linearizability checking \tlaplus (LC).}
\label{tbl:systems}
\begin{tabular}{r|rrrrr}
\thickhline
\textbf{System} & \makecell{\textbf{Impl}\\\textbf{LOC}} &
\makecell{\textbf{Fuzz}\\\textbf{LOC}} &
\makecell{\textbf{\tlaplus}\\\textbf{LOC}} &
\makecell{\textbf{LC}\\\textbf{LOC}} \\
\thickhline
\texttt{WiredTiger} & 129,666 & 614 & 386 & 311 \\
\hline
\texttt{BAT} + \texttt{BAT-EagerDel} & 3,887 & 165 & 57 & 7 \\
\hline
\texttt{ConcurrentQueue} & 2,726 & 174 & 221 & 29 \\
\hline
\texttt{WSDeque} & 102 & 112 & 31 & 11 \\
\thickhline
\end{tabular}
\end{table}

\subsection{Systems Evaluated}
\label{sec:eval-systems}

We evaluated \omnilink on the four systems described below.
Their features are listed in Table~\ref{tbl:systems}.

\subsubsection{WiredTiger}
\label{sec:eval-wt}

WiredTiger~\cite{wiredtiger} is a high-performance, transactional
key-value store, used as the default storage engine for MongoDB. It
organizes data on disk in the form of copy-on-write B+ trees, but after
it loads a tree node to memory, it represents it using a lock-free data
structure based on skip lists. WiredTiger uses multi-version concurrency
control (MVCC) to allow transactions to access different versions of
data concurrently without locking. The combination of these techniques
enables WiredTiger to efficiently handle a high degree of concurrency.
The WiredTiger specification we use~\cite{mongodbStorageTla, schultz2025} models snapshot isolation~\cite{snapshot_isolation95} by individually describing transaction start, commit, abort, and internal operations. 


\subsubsection{Balanced Augmented Tree (BAT)}
\label{sec:eval-bat}

BAT is a highly-scalable concurrent balanced binary search tree which stores key-value 
pairs~\cite{bat}. It efficiently supports range queries, such as counting the number of 
items in a given key range, by augmenting each internal node with the number 
of items in its subtree. 
This is done by maintaining both an immutable tree, which contains augmented values and helps range queries acquire a consistent view of the tree, and a mutable tree, which can be updated more efficiently.
Updates are first applied to the mutable tree. Concurrent updates then collaboratively build a new version of the immutable tree through path-copying and path-merging.
Updates are linearized when this new version of the immutable tree is installed at the root.
All query operations are applied to the immutable tree pointed to by the root.
%

We test two variations of this data structure: BAT and BAT-EagerDel (short for eager delegation). The main 
difference between these is the way that new versions of the immutable tree are built.
BAT forces updates to attempt to build the new version themselves, returning from the update call only 
after the root is sure to have been updated. BAT-EagerDel occasionally delegates 
this task to a concurrent update which allows it to improve performance on 
inserts and deletes. However, this means that BAT and BAT-EagerDel have 
different potential implementation bugs that affect their linearizability. It is worth noting that both types have a formal proof of correctness.

\subsubsection{ConcurrentQueue}
\label{sec:eval-concurrentqueue}

ConcurrentQueue~\cite{cameron314_concurrentqueue} is a widely used C++
library implementing a high-performance, lock-free multi-producer
multi-consumer (MPMC) queue. ConcurrentQueue is representative of
modern concurrent data structures that aggressively
specialize for performance.

ConcurrentQueue builds on techniques from the literature on
non-blocking queues. The implementation uses per-producer segments,
elimination of false sharing, and batched allocation techniques to
achieve high throughput under contention. The numerous optimizations
result in a queue that is fast, but with complex concurrency behavior
that is difficult to reason about manually.
As discussed in Section~\ref{sec:non-lin-queue}, ConcurrentQueue is only substep-linearizable.

\subsubsection{Work Stealing Deque}\label{sec:eval-work-stealing}

We evaluate a linearizable concurrent double-ended queue (deque)~\cite{arora-deque98} crucial for efficient parallel computation in the
parallel algorithm toolkit library ParlayLib~\cite{parlaylib20}. In a queue, elements are inserted at the bottom of the queue and remove
from the top, however a work-stealing deque permits elements to be removed from the top or bottom of the queue. Each thread owns a deque and the
specification enforces that threads may only insert or remove elements from the bottom of the deque they have ownership of. However, a
thread can remove elements from the top of a deque owned by any thread. Operations accessing a deque interact with atomic \emph{top} and
\emph{bot} variables that mark the location of the top and bottom of the deque.

\subsection{RQ1: Bugs Found with \omnilink}
\label{sec:eval-bugs}

\omnilink was able to detect multiple expected and unexpected bugs.
We describe bug detection outcomes for each system we validated.

\subsubsection{WiredTiger Bugs}
\label{sec:eval-bugs-wt}


\omnilink helped the WiredTiger team identify modelling issues in their \tlaplus specifications.
We have confirmed both of these issues with the specification author.

\begin{enumerate}[leftmargin=*, noitemsep, topsep=0pt]
  \item The model incorrectly checked transaction timestamps using $<$
    for validity, while WiredTiger used $\leq$, as it is legal for two
    transactions to modify the same key using the same timestamp.
  \item The model assumed that an application must commit transactions
    in the order of their commit timestamps, which turned out
    to be not true in WiredTiger. WiredTiger allows transactions to commit
    out of order, as long as they do not update the same keys.
\end{enumerate}

While the WiredTiger specification is of a non-linearizable snapshot isolation level storage engine, it is written at a linearizable granularity, so another linearizability checker could have found these bugs.
However, \omnilink made it possible for us to re-use a large existing \tlaplus specification.
This reuse is important since this specification was previously validated by experts using \tlaplus tooling.

\subsubsection{BAT Bugs}

\omnilink was able to detect injected bugs in BAT and BAT-EagerDel, and one bug that was undetected and confirmed by the authors.
All injected bugs were ones encountered by the authors of BAT and BAT-EagerDel during the development of these data structures.
Many were not caught by the authors' test cases and were only discovered initially through careful inspection of the source code (specifically, bugs ~\ref{bug:snapshot-of-child-and-its-version}, ~\ref{bug:unsuccesful-update-inconsistent}, ~\ref{bug:nil-version} below). 

\begin{enumerate}[leftmargin=*, noitemsep, topsep=0pt]
  \item BAT: an injected bug that assigned incorrect child pointers in the immutable tree.
  \label{bug:flipped-child-pointers}
  \item BAT: an injected bug that led to a thread getting an inconsistent view of a node in the mutable tree and its corresponding node in the immutable tree. Describing why this can cause an invalid execution requires a complex counterexample involving 3 threads, 4 operations and an intricate interleaving between them.
  \label{bug:snapshot-of-child-and-its-version}
  \item BAT: an injected bug that caused failed inserts and deletes to return before they were linearized. With this bug, the code still satisfies Sequential Consistency~\cite{lamport1979make}, a weaker correctness condition, but it is not linearizable.
  \label{bug:unsuccesful-update-inconsistent}
  \item BAT: A non-injected bug described below. \label{bug:non-injected}
  \item BAT-EagerDel: an injected bug that affected when a thread that had been delegated work by another thread was allowed to continue propagating augmented values to the root. This bug only rarely triggered a visible error in executions with millions of operations.
  \label{bug:incorrect-delegation}
  \item BAT-EagerDel: an injected bug that incorrectly assigned immutable tree nodes during a balancing rotation of the tree. This bug was only detected originally by a thorough examination of the correctness of the data structure.
  \label{bug:nil-version}
\end{enumerate}

The non-injected bug \ref{bug:non-injected} present in BAT and BAT-EagerDel was in a contains function which checks whether a key is present in the tree.
The implementation of  this function had a mistake where it read values from  an immutable tree not intended to be linearized until later in the execution. 
Detecting a bug of this type without \omnilink is challenging since it does not modify the data structure and still returns a valid result, just one from an invalid timestamp.
By using TraceView to visualize faulty traces, the BAT authors were able to identify a common pattern that both confirmed which function was causing the issue and determined that it was returning a value from a future timestamp, which led to the bug being fixed.

While these bugs can also be found by state of the art linearizability checker Porcupine~\cite{porcupine}, in 42\% of our attempts traces exhibiting these bugs caused Porcupine to run out of memory.
\omnilink's better scalability made it possible to confidently reproduce these bugs.

\subsubsection{ConcurrentQueue}

\omnilink did not find any new bugs in ConcurrentQueue, nor does its issue tracker have any record of concurrency bugfixes we could reintroduce as seeded bugs, despite extensive use.
With the performance-intensive size operation modelled, we were able to evaluate 500 operations across 5 threads.
With that operation disabled, retaining all enqueue and dequeue functionality, we were able to evaluate 500k operations across 5 threads.
Due to ConcurrentQueue being only substep linearizable, \omnilink is the first tool to be able to perform such an analysis.

\subsubsection{Work Stealing Deque Weak Memory Bug}
\label{sec:eval-weak-mem-bug}

Using burst mode, \omnilink was able to find a seeded weak memory bug in an implementation of the work stealing deque~\cite{arora-deque98}.
This bug is representative of a previously fixed implementation error, where memory accesses needed to be set to sequentially consistent, but were configured with more relaxed semantics instead.

The bug can occur when two threads, both of whose deques start empty, concurrently add an element to the bottom of their deque and subsequently
remove an element from the deque owned by the other thread.
When adding an element to the bottom of a deque, a thread increments that deque's \emph{bot} variable and stores the new value. When removing an element from the top of a deque, a thread loads the \emph{bot} and \emph{top} variables for that deque and if they are identical, returns that the deque was empty.
If the store to the \emph{bot} variable is not sequentially consistent, it is possible for both threads' load of the other \emph{bot} to be reordered before this store.
In this case, both removals may return that the deque was empty which is non-linearizable behaviour.

\subsection{RQ2: Expressivity, Implementation, and Specification Effort}
\label{sec:eval-view}

Here we report the manual effort used to validate each system we evaluated.
The sizes of the code we wrote for each part is listed in Table \ref{tbl:systems}.

In general, the process of using \omnilink is:
(1) find or write a draft \tlaplus specification of the system;
(2) implement a randomized workload generator in C++;
(3) gather traces from system executions;
(4) validate traces against \tlaplus model;
(5) diagnose validation errors;
(6) and go to (1) until the acceptable coverage and error-freedom is reached.
For each system we validated, we describe details of the effort required below.

\subsubsection{WiredTiger}
\label{sec:eval-wt-effort}
WiredTiger is unique in this list, in that we validated it against an existing \tlaplus specification~\cite{mongodbStorageTla, schultz2025} named \mpcal|Storage|.
To make use of the \mpcal|Storage| specification, we needed to adapt it to the form expected by \omnilink.
In total, we wrote 265 lines of additional \tlaplus, and made a 2 line change to the original specification.

We spent a person day writing adaptor actions to drive the original \tlaplus.
For instance, we needed to map between the model's representation of success/error conditions and API error codes.

To exercise the system, we wrote 590 lines of C++ code that logged randomized calls to the WiredTiger API.
The main effort needed was to work around assertion failures, where the API would reject certain calls by terminating the whole process.
Refining the runner took 1.5 person weeks.

Given the work above, the system could be validated end to end, but it did not scale.
This was a consequence of how the \mpcal|Storage| model was written: it tracks all transactions and log entries in existence without bound.
As a result, executions beyond 1,000 actions resulted in large logs and transaction sets copied per state, slowing down \tlaplus next-state relation evaluation.
Normal model checking will focus on permutations within a short log, rather than linearization over a long log, so it is normal for \tlaplus specifications to have this type of issue.

In a conventional linearizability checker, we would have had limited recourse for getting better performance, since no known optimizations target state size.
With \omnilink, we used its substep feature.
For each action that could increase log length (transaction prepare, commit, rollback to stable), we defined the original action in \mpcal|Storage| as the first (non-final) substep.
Then, we added a log compaction substep.
The log compaction substep's job is to opportunistically remove and merge ``old'' log elements.
We found that only log elements older than both the oldest active transaction and the ``oldest timestamp'' value could be soundly compacted.
50 of the lines of \tlaplus we report in Table \ref{tbl:systems} are ``log compaction'' actions we developed over 4 person days.

Interestingly, the entire log compaction issue could not be solved in \tlaplus.
It is easy to randomly generate a sequence of actions that force retention of an impractically large log.
Part of our solution was to enforce that the ``advance oldest timestamp'' action was chosen at least every 20 commits of the test workload, allowing regular compaction of the modeled log. 

\subsubsection{BAT}
We validated BAT against a purpose-written \tlaplus specification.
BAT offers strong linearization guarantees, so the main effort was to precisely model API edge cases, such as number ranges being inclusive or exclusive.
We quickly exposed most bugs with \omnilink, and the process of building the validation setup took three person days, during which the \tlaplus specification and fuzzer quickly stabilized.


For BAT, diagnosing \omnilink's validation failures was the most challenging part, requiring careful analysis using the TraceView utility.
We enabled identity path pruning for read-only operations like ``contains'' and ``range query'', but no further performance engineering was necessary.

\subsubsection{ConcurrentQueue}

Our \tlaplus specification of ConcurrentQueue was based on its formalization in \cite{hackett2026substep}, where the queue is represented as a collection of thread-local queues.
Some operations target a specific thread-local queue, and some operations non-deterministically scan all queues.
Writing the specification was only possible using substeps, as many of the operations non-linearizably check multiple sub-queues while interleaving with other operations.
Thanks to iterating with \omnilink, we were able to reject multiple early \tlaplus specifications drafts after seeing counterexamples.
Overall, \tlaplus specification and C++ development took 14 person days.

The main challenge was performance engineering.
Checking substeps requires more work, since each substep has a similar performance effect as checking an individual operation.
Queues also allow long sequences of enqueues, during which there may be no evidence of whether values are dequeued in the correct order.
This requires any linearizability checker to speculate over long subsequences of operations.

In the end, we were unable to leverage any known linearizability checking optimizations, though we were able to validate traces with up to 500k operations.

\subsubsection{Work Stealing Queue}\label{sec:eval-work-stealing-effort}

It took less than 1 person day to model operations on the Work Stealing Queue, partly because they have the same form as ConcurrentQueue with which we had experience, and the operations are simpler with no sub-step behavior.
The main effort was in preventing logging from interfering with weak memory effects.
Surprisingly, writing to a thread-local byte array still prevented reproducing our seeded bug.
In the end, it took 2.5 person days to isolate the effect, at which point \omnilink was able to read the partial trace and detect the bug.

\subsection{RQ3: Log Vector Optimization Impact}
\label{sec:eval-state-of-the-art}

\begin{figure*}
  \begin{minipage}{0.64\textwidth}
    \begin{minipage}[t]{0.48\textwidth}    
    \includegraphics[width=\textwidth]{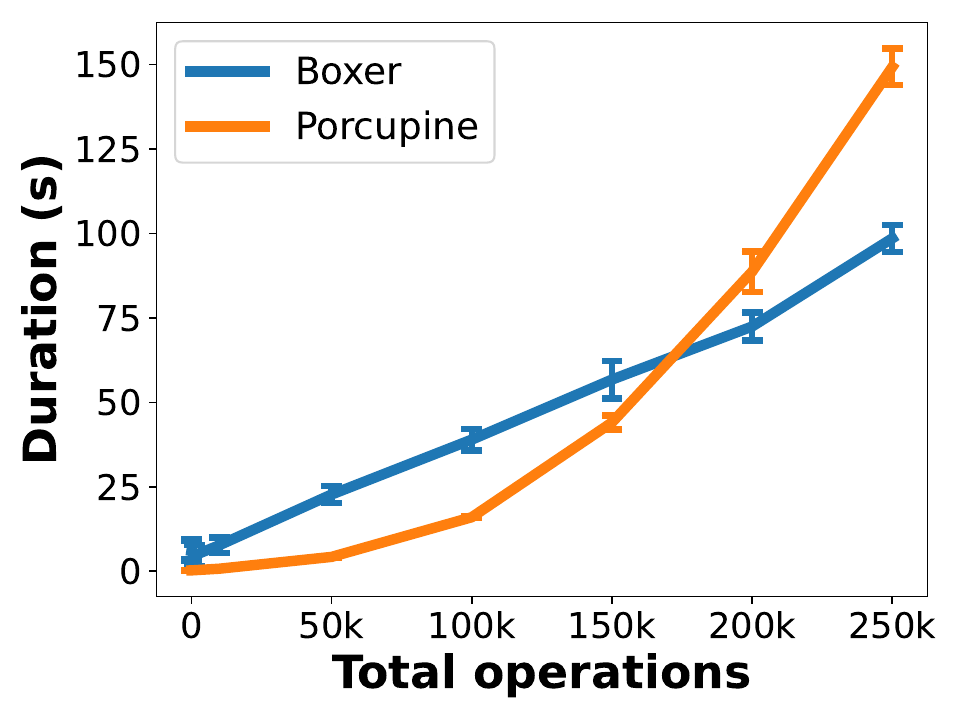}
    \subcaption{Runtime in seconds, varying length of trace.\label{fig:porcupine-time}}
    \end{minipage}\hfill
    \begin{minipage}[t]{0.48\textwidth}    
    \includegraphics[width=\textwidth]{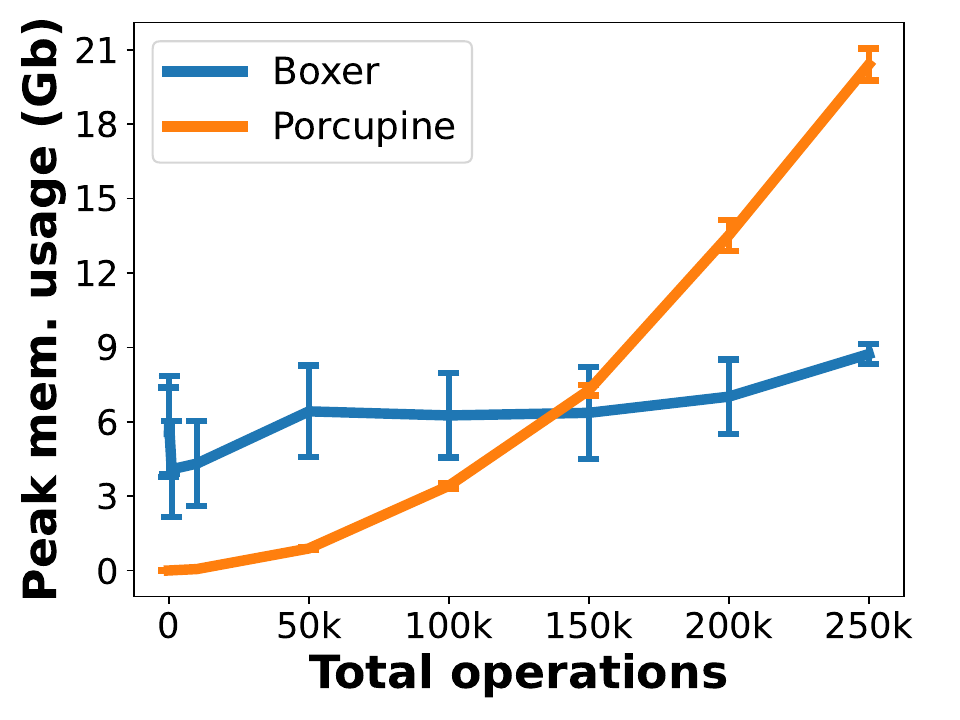}
    \subcaption{Peak memory usage in GB, varying length of trace.\label{fig:porcupine-mem}}
    \end{minipage}
    \\
    \begin{minipage}[t]{0.48\textwidth}    
    \includegraphics[width=\textwidth]{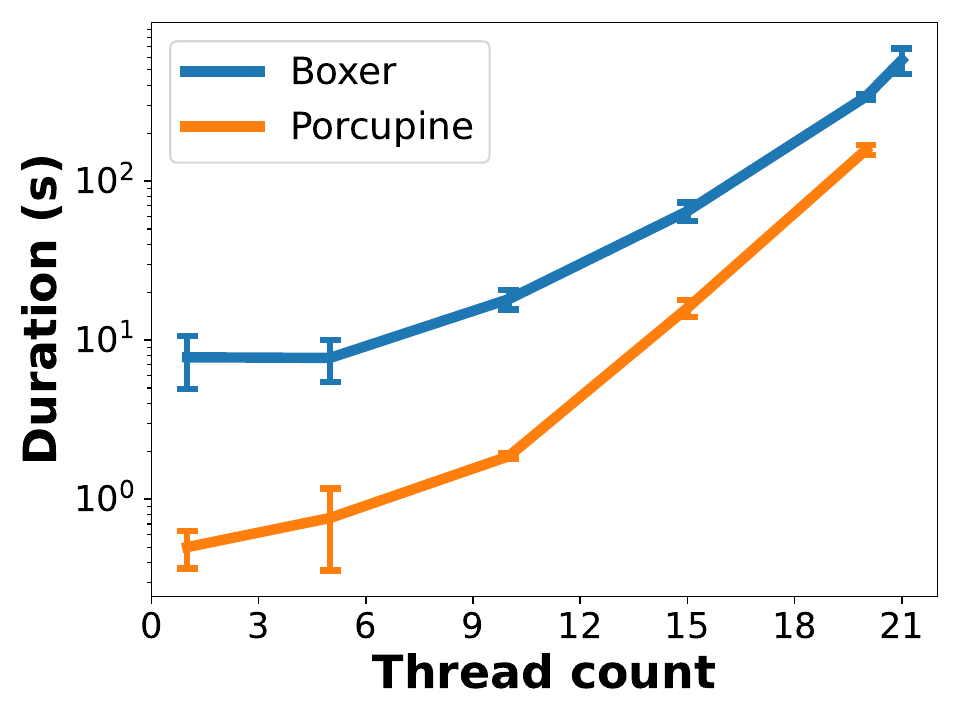}
    \subcaption{Runtime in seconds, varying thread count.\label{fig:porcupine-time-threads}}
    \end{minipage}\hfill
    \begin{minipage}[t]{0.48\textwidth}    
    \includegraphics[width=\textwidth]{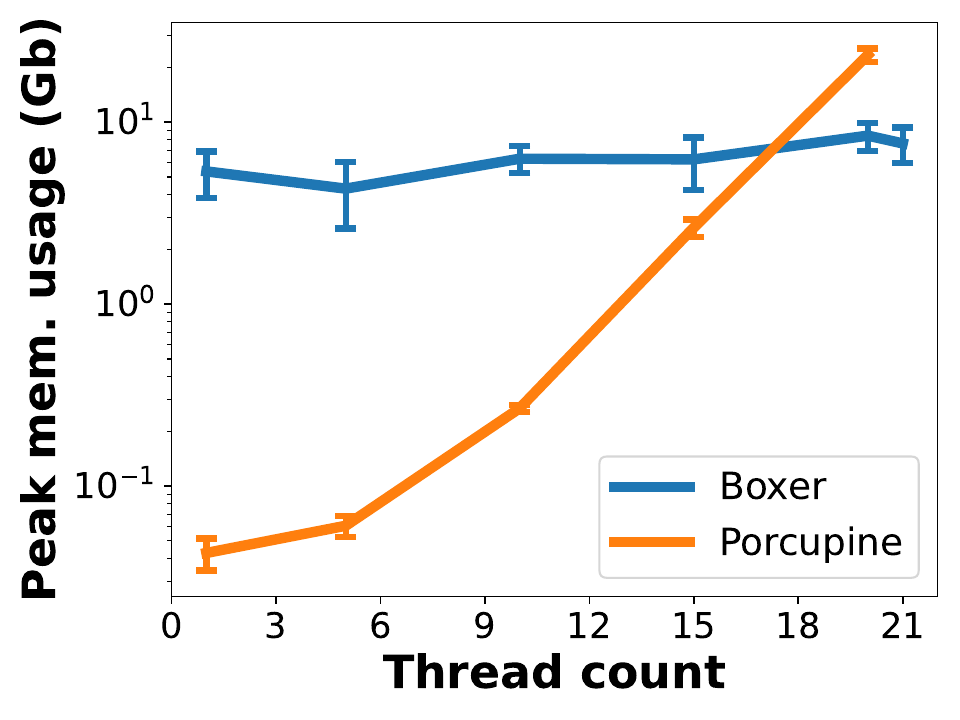}
    \subcaption{Peak memory usage in GB, varying thread count.\label{fig:porcupine-mem-thread}}
    \end{minipage}
\caption{Peak main memory and runtime of Porcupine and \omnilink validations for accepted(valid) traces of different lengths. When varying total operation count, thread count is fixed at 5. When varying thread count, total operation count is fixed at 10K.}\label{fig:porcupine-all}
\end{minipage}
\hfill
\begin{minipage}{0.32\textwidth}
\begin{minipage}[t]{0.96\textwidth}    
    \includegraphics[width=\textwidth]{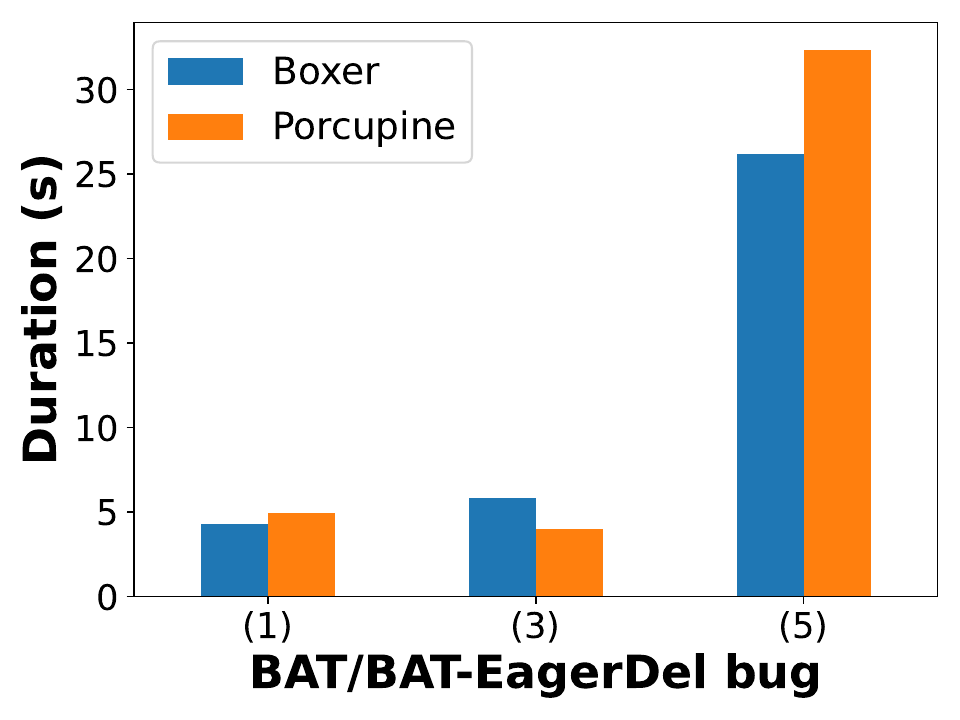}
    \subcaption{Detection speed of various BAT bugs in seconds, <30s detection time.}
    \end{minipage}
    \\
    \begin{minipage}[t]{0.96\textwidth}    
    \includegraphics[width=\textwidth]{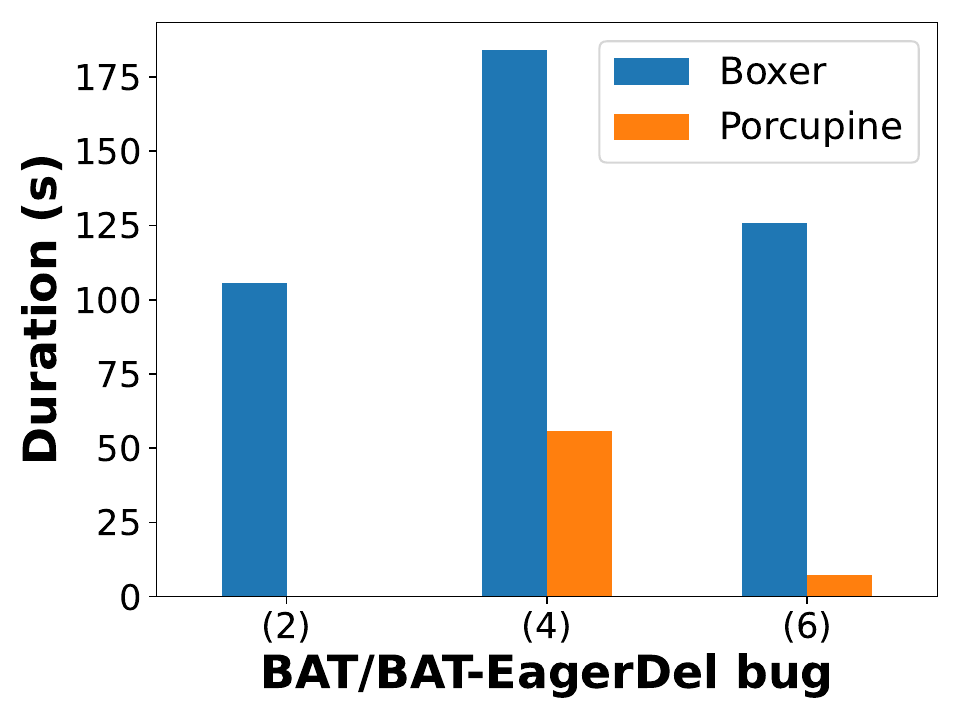}
    \subcaption{Detection speed of various BAT bugs in seconds, >30s detection time.}
    \end{minipage}
\caption{Runtime of Porcupine and \omnilink validations for rejected (invalid) traces of various BAT bugs. Thread count is fixed at 5.\label{fig:porcupine-failing}}
\end{minipage}
  
\end{figure*}

We compare \omnilink with Porcupine~\cite{porcupine} and VeriLin~\cite{jia2023verilin}.
Porcupine is a Go implementation of a state of the art linearizability checking routine~\cite{horn2015}.
VeriLin is a linearizability checker for Java data structures with novel design features for scalability.
Note that neither Porcupine nor VeriLin have support for substep linearizability.

We gathered traces from BAT into \omnilink's database, including all known buggy versions, and then validated them using \omnilink, Porcupine, and VeriLin.
To compare the tools fairly, we translated our \tlaplus model of BAT into Go for Porcupine and Java for VeriLin.
We did not use WiredTiger or ConcurrentQueue, as WiredTiger is labor intensive to translate, and ConcurrentQueue is only substep-linearizable.
For each translation, we wrote a loader from \omnilink's trace database into the other verifiers' in-memory format.
The translations were 1--1, and the tool outputs always agreed with one another.

While VeriLin was able to correctly validate small logs up to 100 actions, it timed out in all other tests, and we exclude it from our graphs for that reason.
We are confident that this is because it relies on an unrealistic requirement for timebox distribution: their design assumes that timeboxes always form small, disjoint cliques.
The system executions we recorded were optimized to pack timeboxes together as densely as possible and find the limits of the tools under test, which broke VeriLin's design assumption.

For \omnilink and Porcupine, we recorded data series varying workload thread count and operation count, measuring wall clock run time and memory usage during validation by each tool.
We captured 5 different logs for each pair of thread count and operation count, and validation was run once per log.
We also captured 5 repetitions of buggy traces for all 6 bugs we describe for BAT, in that case fixing thread count at 5 and operation count at 200k.
Each validation was given a ceiling of 32GB memory usage using benchexec~\cite{benchexec}, and validation was terminated if the ceiling was reached.

\subsubsection{Varying Operation Count}

Figure~\ref{fig:porcupine-time} compares the tools' scalability by measuring wall clock runtime, increasing total operation count from 5 to 250K while maintaining a thread count of 5.
We can see that Porcupine performs better for small to medium trace sizes, while \omnilink scales better as the operation count increases.
This is because of \omnilink's log vector optimization, which allows \omnilink to track which operations have been linearized without repeatedly copying an $N$-bit bit vector $N$ times, where $N$ is the number of operations in the trace.

Figure~\ref{fig:porcupine-mem} shows the same experiment as Figure~\ref{fig:porcupine-time}, but tracking peak memory consumption instead of runtime.
Porcupine consumes the least memory for small workloads, but its memory consumption increases exponentially with log size due to its factor of $N$ memory overhead as $N$ increases.
\omnilink also consumes memory proportional to the state space it covers, but each state it discovers only costs the state's storage footprint and as many 32-bit integers as there are threads, which lets it scale more efficiently.
To rule out the possibility that \omnilink's disk cache is hiding its memory usage, note that \omnilink's total disk footprint after running all experiments (with no deletion but storing every visited state just once) is 22GB --- less than one Porcupine run with a large log --- which suggests an algorithmic improvement in memory efficiency.

\subsubsection{Varying Thread Count}

Figures \ref{fig:porcupine-time-threads} and \ref{fig:porcupine-mem-thread} vary thread count from 1 to 21 for a constant 10k operations, to show how the tools scale for larger numbers of threads.
Thread count is a different challenge than operation count, because, given a densely packed set of timeboxes, more threads increases the number of timeboxes that may overlap at once.
A linearizability checker must enumerate all interleavings of overlapping timeboxes, so for $n$ threads, the number of combinations across $n$ concurrent actions grows as $n!$ in the worst case.
Looking at wall clock time, both tools show a similar curve, with Porcupine performing better overall but with faster growing memory consumption.
The wall clock result is coherent with other results given 10k operations, while the memory result is a consequence of every evaluated interleaving storing another copy of Porcupine's large bit vector.
At 21 threads, Porcupine consistently reaches the memory threshold of 32GB per experiment and terminates abnormally, while \omnilink still finishes after over 15 minutes.
This experiment shows that neither tool is practical for large groups of concurrent threads, though \omnilink has the potential to scale further if the fundamental state space blow-up can be mitigated.

\subsubsection{Time to Detect Known Bugs}

Figure~\ref{fig:porcupine-failing} shows how long \omnilink and Porcupine take to detect each known bug in the BAT systems, given 200k operations and 5 threads per experiment.
Note that while Porcupine never disagrees with \omnilink and is often faster, it cannot always detect these bugs.
In 42\% of cases where \omnilink detects a bug, Porcupine runs out of memory.
For bug (2), it failed in all cases.

We confirmed that a log size of 200k operations is necessary to reliably detect these bugs.
The average length of the counterexample \omnilink reports is 199k $\pm$ 58k, so choosing a shorter log where Porcupine performs better will correspondingly decrease the ability to detect these bugs.
For 200k operations the bugs are reached in 29/30 experiments, while for 100k operations only 21/30 executions exhibit the intended bug, meaning the same experiment could potentially miss 2/6 bugs (if misses were clustered around specific categories).
This demonstrates that our log vector optimization meaningfully improves the bug-finding usefulness of linearizability checking tools.

\subsection{RQ4: Impact of Path Pruning}
\label{sec:eval-path-pruning}

To measure the impact of path pruning on \omnilink's systems, we validated the same 5 traces per system, enabling and disabling path pruning.
To defeat \omnilink's ability to cache state transitions across similar traces (in this case, the same one), we loaded a database with only the log and no states explored before each run.

Overall, we find path pruning is applicable to the BAT system, but there are difficulties in applying it to large specifications, or in cases where a point-wise optimization like path pruning does not exist.
In WiredTiger, while prunings were possible, the overhead of evaluating their \tlaplus definitions exceeded their benefit.
In part, this motivates exploring more flexible \tlaplus evaluation and path pruning definition forms.
More fundamentally, we leave exploring optimizations that allow lookahead, such as an adaptation of DPOR techniques, as future work.

\textbf{BAT.}
We used buggy traces with 100k operations and 5 threads, and we used identity path pruning to optimize read-only operations.
On average, path pruning reduced the explored state count by $1.75\times$, and reduced wall clock time by $1.2\times$.
Running the same experiment with valid traces, there was a benefit of only $1.09\times$ decrease in validation time.
Note that path pruning will have an outsize effect on validation failures, since they require exhausting all possible paths, while successes only need to find one linearizable path.

\textbf{WiredTiger.}
We used valid traces with 50k operations and 5 threads.
Our path pruning optimization was able to prune between 4 and 29 states total per validation, maintaining a 36s average validation time.
We attribute the limited impact of path pruning here to their limited applicability.
While the log compaction step we describe in previous sections is by definition suitable for generalized identity path pruning, most parts of the WiredTiger transaction lifecycle have no sound path pruning.
That said, performance with path pruning disabled was sufficient to find the two specification bugs we report.

\textbf{ConcurrentQueue.}
We were not able to apply any known path pruning optimizations to ConcurrentQueue.
The full model of ConcurrentQueue includes an approximate size check, where each sub-queue is measured individually using sub-steps and the sum of all the sizes is reported.
The substeps due to the size check operation have effects on state since they accumulate partial observations for future substeps, so they are not subject to identity pruning.
Different permutations of enqueues and dequeues affect which possible approximate sizes are visible, so we did not find any generalized identity path prunings.
The dequeue and size operations implicitly scan all sub-queues, so P-Compositionality is inapplicable.
It is also possible to implicitly measure that the total queue size was zero if a dequeue operations returns empty, so even if we rule out the size operation, dequeues alone can cause all the problems discussed above.

\textbf{Work Stealing Queue.}
Each operation on the work stealing queue identifies a specific target queue and affects only that queue, so partitioning the log using P-Compositionality is possible.
While we prove that \omnilink can perform this optimization in Section~\ref{sec:p-comp-path-pruning}, a non-fundameWe were able to validate traces with up to 5k operations and 5 threads.ntal implementation limitation prevents evaluating it.
Specifically, the interface for writing path pruning definitions cannot currently refer to timeboxes; it can only use state-level information.

\subsection{RQ5: Effectiveness on Weak Memory with Burst Mode}

We show that \omnilink can detect issues caused by weak memory semantics by using it to detect a seeded bug in the Work Stealing Queue system.
We ran the system under test with two threads and 200k operations total, and recorded 5 logs.
\omnilink could detect a correctness violation in all 5 logs at an average depth of 110k $\pm$ 61k operations, and when the problematic memory access was adjusted no violations could be found.

As we describe in Section~\ref{sec:eval-work-stealing-effort}, logging while allowing weak memory effects between operations is distinct from the usual method of measuring timestamps at each invocation and response.
Minor changes had to be made to our default logging system to accommodate this, and we leave a more robust logging method as future work.

Once the missing timeboxes are filled in, burst mode traces can in principle be checked by any linearizability checker that tolerates events on the same thread with overlapping timeboxes.
But, we observed that existing linearizability checkers treat overlapping operations by the same thread as concurrent with one another, and therefore reorderable.
Such a reordering is not sound.
This happens with Porcupine, meaning it cannot soundly validate burst mode logs. 

\subsection{Concurrency Testing with Fray and \omnilink}
\label{sec:fray}

Concurrency testing frameworks like Fray~\cite{li2025fray} are
effective at generating diverse thread interleavings, but they
typically rely on simple correctness oracles such as assertion
failures, uncaught exceptions, or crashes. By integrating \omnilink
with such frameworks, we can both (1) validate a broader range of
concurrent executions than those covered by a fixed test suite, and
(2) check those executions against richer semantic specifications
expressed in TLA+.

As a case study, we integrated \omnilink with Fray and applied it to
Java's \texttt{ConcurrentMap}. We extended a TLA+ map specification
covering the basic operations \texttt{put}, \texttt{remove},
and \texttt{containsKey} to additionally model the \texttt{size()}
operation. Prior work~\cite{concurrent-size-oopsla22} has shown
that \texttt{size()} does not satisfy linearizable semantics in common
implementations, yet this behavior is difficult to detect using
conventional assertions.

We integrated \omnilink with Fray so that executions generated during
concurrency testing are exported as traces and then validated against
the TLA+ specification. Using this integration, we
exercised \texttt{ConcurrentMap} operations generated by Fray and
identified executions that violate a linearizable map specification
when \texttt{size()} is included. This demonstrates that \omnilink can
augment concurrency testing frameworks with specification-based
validation, enabling the detection of semantic consistency violations
that are not naturally exposed by conventional assertions or
crash-based checks.




\subsection{Implementation Debugging with rr}
\label{sec:rr}

Given a counterexample from \omnilink, we are able to
explore the corresponding implementation behavior using rr's replay
feature~\cite{rr-project}.  For every trace we record for \omnilink, we record the
concrete implementation behavior in parallel using rr's whole-program
recording capability.  We can choose at any time to unpack this
recording and replay the concrete execution. This is useful as it
makes \omnilink's randomized fuzzy behavior deterministically
replayable.

Since \omnilink's logging format includes precise implementation
timestamps, we are able to insert conditional breakpoints that pause
the replay at the instruction where an operation's starting timestamp
is recorded.  That breakpoint will only pause the program if the
timestamp matches the specific action we are interested in.  Since rr
deterministically replays all syscalls and threading decisions, all
randomized behavior and timestamp measurements play out identically.
Therefore, our breakpoint will allow us to replay specifically the
action we are interested in, as it originally happened.  We can then
single-step from that point and examine any information accessible via
conventional debugging, including information that was not present in
the original trace.

We have used this technique several times during \omnilink's development and it works on all the systems we evaluate (except for reproducing weak memory effects).
The main difficulty can be preventing timestamp-measuring code from being optimized by the compiler such that it is not possible to breakpoint, which can be avoided with sufficient care.

\section{Related Work}
\label{sec:rw}



\textbf{Trace checking:} 
Pick et al.~\cite{pick2025checking} developed a trace checking technique for verifying isolation guarantees provided by database systems.
Elle~\cite{kingsbury2020elle} also examines traces for potential defects, but is exclusive to database transactions. Niu et al. ~\cite{zookeepervalidation2022} validate distributed system traces against \tlaplus specs, focusing on ZooKeeper. Their approach depends on hand-instrumented internal logs and custom translators. Trace checking has been attempted for large commercial codebases, but has been found to be impractical~\cite{Davis2020Extreme} or has required months of effort by a large team~\cite{howard2024smartcasualverificationconfidential}.
\omnilink is not limited to checking database transactions and requires lighter-weight instrumentation, with more options to infer abstracted behavior.
With TraceLink~\cite{hackett_2025_tracelink}, engineers write a spec in Modular PlusCal and transpile it to Go with PGo, which automatically adds instrumentation to record events along with their vector clocks.
\omnilink treats the implementation as a black box, and uses timeboxing instead of vector clocks to resolve ambiguity.


\textbf{Linearizability testing} is a special case of trace checking for concurrent data structures. 
In general, given a trace and a sequential specification, testing if the trace is linearizable is NP-hard~\cite{gibbons1997testing}.
Many 
search approaches have been proposed for linearizability testing~\cite{winggong1993,lowe2016history,ozkan2019hittingfamilies,vechev2009spin, horn2015faster,koval2023lincheck,porcupine}, some for weak memory models~\cite{relinche2025,cdsspec2017}. 
Lineup~\cite{burckhardt2010line} proposes automatically inferring the sequential specification by running the concurrent data structure on one thread.
For specific abstract data types such as sets, queues and stacks, more efficient algorithms have been developed that run in polynomial time~\cite{abdulla2025efficient,peterson2021concurrent,emmi2017sound,han2024efficient}.
For example, testing linearizability of sets supporting insert, delete, and lookup can be done in $O(n)$ time where $n$ is the number of operations in the trace, irrespective thread count~\cite{abdulla2025efficient}.
Importantly, existing polynomial time algorithms do not work for more advanced set data structures that support range queries such as the balanced augmented tree~\cite{bat} that we validated with \omnilink.

\textbf{Substep linearizability:} Various generalizations and alternate definitions of linearizability have been proposed~\cite{garg1999normality,gil1994setlin,goubault2019generalizedlin,hemed2015calin,haas2016locallin}. Casta\~{n}eda et al.~\cite{cataneda2018intervallin} introduce interval-linearizability where operations are assigned linearization intervals rather than linearization points. Vale et al.~\cite{vale2023complin} propose an abstract definition of linearizability equivalent to interval-sequential linearizability up to removal of pending invocations. We have found the behavior of objects like ConcurrentQueue easier to model with substep linearizability than with these alternate definitions.

\textbf{Runtime verification (RV):} In RV,
a monitor continuously checks if the implementation conforms to the spec~\cite{rv_dist_intro_2018}.
For example, Choreographic PlusCal~\cite{Foo2023ChoreoPCal} compiles a global protocol spec into local runtime monitors for each participant.
This permits fine-grained conformance checking, but requires modifying the application to embed monitors.
Ellsberg~\cite{ellsberg} relies solely on observation of participants' RPCs to continuously check their conformance.
Ellsberg requires an engineer to rewrite the spec in the Ellsberg language.

\textbf{Test case generation:} 
Several works use a \tlaplus spec to generate test cases~\cite{multigrained, sandtable, Wang2023Mocket, Gao2025iMocket}, but require manual changes to the implementation.
%
%
Unlike these systems, \omnilink does not need to interfere with implementation execution. It infers the system's state entirely from an opaque API. Other works compute an approximate minimal set of tests to cover the spec's state graph~\cite{modularVerification, kayfabetlaconf2020}.
Several systems~\cite{model-guided-fuzzing-oopsla25,trace_aware_testing_2019} use the spec to guide a fuzzer, probabilistically optimizing coverage. 
Other fuzzing techniques have been developed for generating schedules that are more likely to produce bugs~\cite{li2025fray,musuvathi2008finding}.
\omnilink relies on existing tests to produce traces, which it checks for conformance. 
\section{Conclusion}


We introduced \omnilink, a methodology for validating traces of
unmodified concurrent systems against \tlaplus specifications using
substep linearizability and user-customizable path pruning optimizations.
By combining reusable formal specifications with scalable
trace validation, \omnilink makes it practical to check rich
correctness properties of concurrent implementations without modifying
the systems under test. Our results demonstrate that this approach can
efficiently detect both injected and previously unknown bugs in
real-world concurrent systems.



\newpage

\bibliographystyle{ACM-Reference-Format}
\bibliography{paper}

\appendix


\section{Understanding \omnilink's Output}
\label{sec:understanding}


Because \omnilink's validation process is based on linearizability solving, interpreting validation failures poses distinct challenges.
When TLC detects a violation, it emits a counterexample representing one of the longest non-failing paths through the search space.
This is a good start, but it does not show us other candidate linearizations, nor does it provide a way to visually inspect the timeboxes of actions surrounding the failure point.
TLC also does not allow us to inspect why an action was not taken without invoking a dedicated debug session.

To make triaging \omnilink outputs practical, we develop a new viewer called TraceView,

Although we leave usability evaluation for future work, TraceView provides a practical means of exploring \omnilink counterexamples.
In our evaluation, we extensively relied on the tool to triage and interpret buggy traces.

\begin{figure*}
    \begin{minipage}{0.96\textwidth}
    \vspace{-18mm}
        \includegraphics[width=\textwidth]{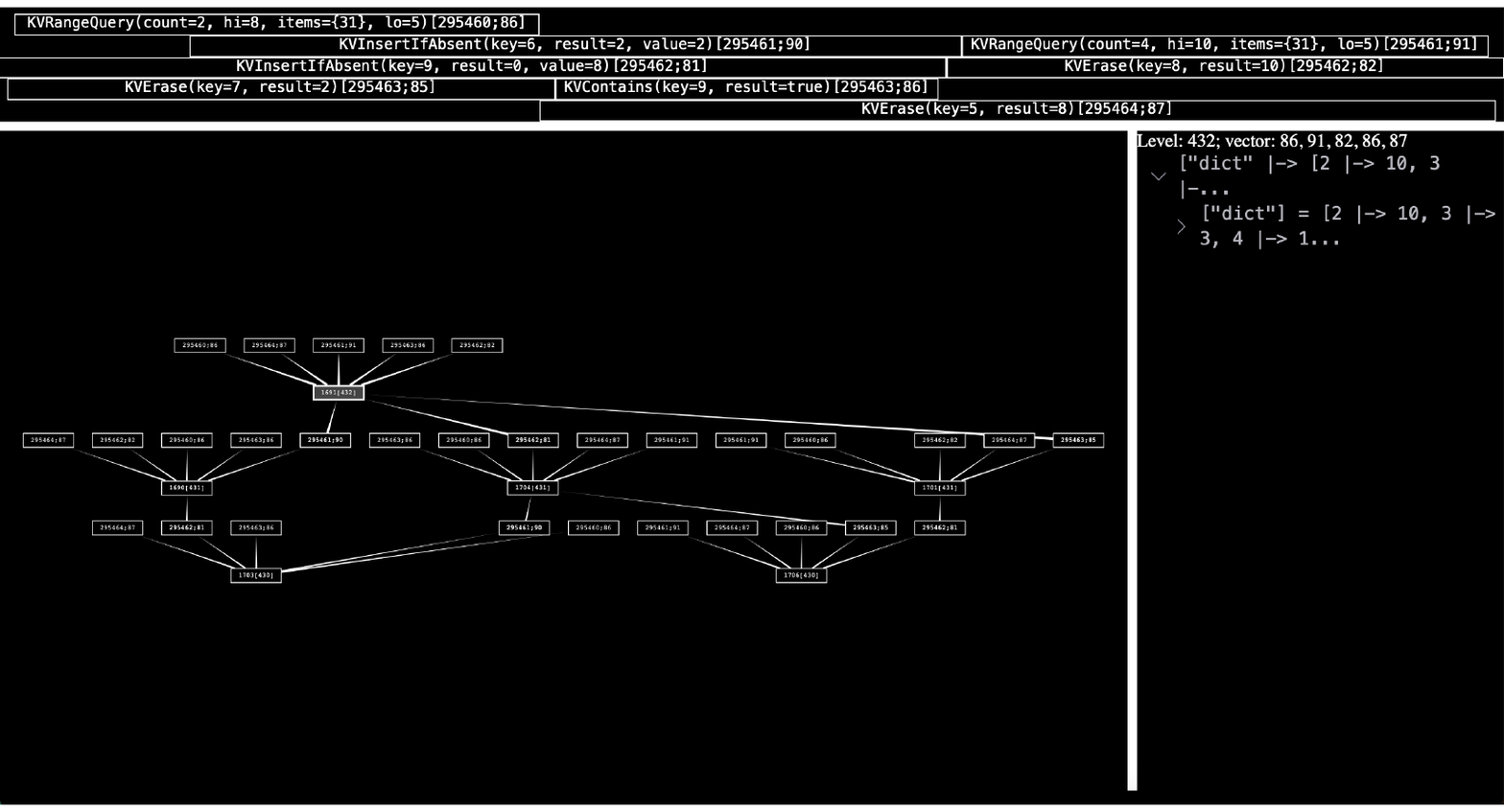}
        \vspace*{-22mm}
    \caption{Using TraceView to visualize a buggy BAT trace. The top of the image shows TraceView's raw timebox view of executed operations and how they overlap. On the left is a graphical representation of visited states and possible state transitions along with the corresponding operation used for the transition. At the topmost state, there are no possible transitions for the algorithm to take. On the right is a display of information about the currently selected state.}\label{fig:traceview}
    \end{minipage}
\end{figure*}

\paragraph{Exploring Failures with TraceView}

Existing \tlaplus tools~\cite{tlatoolbox2019,spectacle} are limited in handling the long implementation-driven traces characteristic of trace validation.
Unusually for TLC outputs, \omnilink might solve around 10 counterexample traces, but each might contain over 100k actions.
Note that prior trace validation work~\cite{howard2024smartcasualverificationconfidential,cirstea:validating} recommends using the TLC Debugger interactively, but this approach provides no summary view, and does not scale to \omnilink's more complex counterexamples.

To visualize these unique model checking outputs, we created \emph{TraceView}, a lightweight visualizer tailored to \omnilink's output. 
TraceView operates on previously computed counterexamples, implementation traces, and state spaces.
Its key feature is \emph{reverse reasoning}: it begins from the
\emph{end} of a counterexample and allows users to explore predecessor
states on demand.\footnote{Note that all other \tlaplus tools we know about start from the initial state.}
Typically, a linearization failure can be diagnosed using only states near the point where linearization failed, so this is in line with how a user expects to diagnose issues.
Rendering from the end state also avoids rendering the expected >100k prior states \omnilink might output, which overloads other tooling. Figure~\ref{fig:traceview} shows an example of the TraceView interface.

Also unlike other counterexample displays, TraceView shows all counterexamples at once using a state graph, allowing a user to easily examine the true set of all conflicting interpretations up to linearization failure, and then expand states backwards as needed.
Operations that were candidates for linearization are displayed as boxes, either dangling if not taken, or linking two states if they were used at the point during linearization.
This enables exploration of deeply ambiguous counterexamples while keeping the visualization manageable.

A concern unique to \omnilink's linearizability solving is the need to view the timeboxes involved in a linearization failure.
Which events do or do not overlap, as well as their order in real time, is essential to accurate debugging.
As a result, TraceView also provides a raw timebox view.
Focusing a timebox by clicking it highlights where it is (potentially in multiple positions) in the state graph.

\end{document}